\documentclass[twocolumn,twocolappendix]{aastex631}
\usepackage{amsmath, amssymb}
\usepackage{graphicx} \usepackage{footnote}
\usepackage{url}

\usepackage{soul} \usepackage{xcolor}
\usepackage{comment}

\usepackage{hyperref}

\newcommand\soutpars[1]{\let\helpcmd\sout\parhelp#1\par\relax\relax}

\newcommand{\rd}{\mathrm{d}}
\newcommand{\rD}{\mathrm{D}}
\newcommand{\rw}{\mathrm{w}}
\newcommand{\rc}{\mathrm{c}}
\newcommand{\rr}{\mathrm{r}}
\newcommand{\rt}{\mathrm{t}}
\newcommand{\ra}{\mathrm{a}}

\newcommand{\rh}{\mathrm{h}}

\newcommand{\brw}{\mathbf{w}}
\newcommand{\br}{\mathbf{r}}

\newcommand{\bL}{\mathbf{L}}
\newcommand{\deltaD}{\delta_{\rD}}
\newcommand{\Lz}{L_z}

\newcommand{\brR}{\mathbf{R}}
\newcommand{\brI}{\mathbf{I}}

\newcommand{\p}{\partial}

\newcommand{\trh}{t_{\mathrm{rh}}}

\newcommand{\rHU}{\mathrm{HU}}

\newcommand{\bw}{\boldsymbol{w}}
\newcommand{\bbv}{\boldsymbol{v}}

\newcommand{\vrr}{v_{\rr}}
\newcommand{\vrt}{v_{\rt}}

\newcommand{\bt}{\mathbf{t}}

\newcommand{\hw}{\hat{\bw}}
\newcommand{\hrt}{\hat{\bt}}

\newcommand{\Rc}{R_{\rc}}

\newcommand{\rra}{r_{\ra}}

\newcommand{\rrh}{r_{\rh}}

\newcommand{\HG}{{}_{2}F_{1}}
\newcommand{\half}{\tfrac{1}{2}}
\newcommand{\Eo}{E_{0}}
\newcommand{\Lo}{L_{0}}
\newcommand{\mH}{\mathbb{H}}
\newcommand{\Ftot}{F_{\mathrm{tot}}}
\newcommand{\tE}{\widetilde{E}}
\newcommand{\tL}{\widetilde{L}}

\newcommand{\stod}{Stod\'o\l{}kiewicz}
\newcommand{\krios}{{\tt KRIOS}}
\newcommand{\raga}{{\tt RAGA}}

\defcitealias{giesers2019}{G19}
\defcitealias{Farrah2023a}{F23}
\defcitealias{CB73}{CB73}
\defcitealias{Zhao1996}{Z96}
\defcitealias{HO1992}{HO92}

\begin{document}

\title{\krios: A new basis-expansion $N$-body code for collisional stellar dynamics}

\correspondingauthor{Kerwann Tep}
\email{tep@unc.edu}

\author[0009-0002-8012-4048]{Kerwann Tep}
\affiliation{Department of Physics and Astronomy,
    University of North Carolina at Chapel Hill,
    120 E. Cameron Ave, Chapel Hill, NC, 27599, USA
}

\author[0000-0003-0341-6928]{Brian T. Cook}
\affiliation{Department of Physics and Astronomy,
    University of North Carolina at Chapel Hill,
    120 E. Cameron Ave, Chapel Hill, NC, 27599, USA
}

\author[0000-0003-4175-8881]{Carl L.~Rodriguez}
\affiliation{Department of Physics and Astronomy,
    University of North Carolina at Chapel Hill,
    120 E. Cameron Ave, Chapel Hill, NC, 27599, USA
}

\author{Jiya Jolly}
\affiliation{Department of Physics and Astronomy,
    University of North Carolina at Chapel Hill,
    120 E. Cameron Ave, Chapel Hill, NC, 27599, USA
}

\author{Eddie Sawin}
\affiliation{Department of Physics and Astronomy,
    University of North Carolina at Chapel Hill,
    120 E. Cameron Ave, Chapel Hill, NC, 27599, USA
}

\author[0000-0003-1517-3935]{Michael S.~Petersen}
\affiliation{Institute for Astronomy,
 University of Edinburgh, Royal Observatory,
 Blackford Hill, Edinburgh EH9 3HJ, UK
}

\author{Christoph Gaffud}
\affiliation{Department of Physics, 
Carnegie Mellon University, 
5000 Forbes Ave, Pittsburgh, PA, 15213, USA
}

\begin{abstract}

The gravitational $N$-body problem is a nearly universal problem in astrophysics which, despite its deceptive simplicity, still presents a significant 
computational challenge.  For collisional systems such as dense star clusters, the need to resolve individual encounters between $N$ stars makes the direct 
summation of forces -- with quadratic complexity -- almost infeasible for systems with $N\gtrsim 10^6$ particles over many relaxation times.  At the same time, the most 
common Monte Carlo $N$-body algorithm -- that of H\'enon -- assumes the cluster to be spherically symmetric. This greatly limits the study of many important 
features of star clusters, including triaxiality, rotation, and the production of tidal debris. 
In this paper, we present a new hybrid code, \krios, that combines 3D collisionless relaxation using an adaptive self-consistent field method with collisional 
dynamics handled via H\'enon's method. We demonstrate that \krios\ can accurately model the long-term evolution of clusters and provide its complete 
phase-space information over many relaxation times.  As a test of our new code, we present detailed comparisons to well-known results from stellar dynamics: (i) the 
collisional evolution of a family of Plummer spheres with varying anisotropy and rotation to core collapse, and (ii) the emergence of the radial-orbit 
instability in radially anisotropic star clusters, including its non-spherical effects.

\end{abstract}


\section{Introduction}

Modeling self-gravitating systems and their long-term evolution has been an active area of research for more than half a century. For large systems such as galaxies and cosmological volumes, modern simulations are now regularly performed with $N> 10^{10}$ particles
\cite[e.g.,][]{Springel2018,Maksimova2021}.  In these ``collisionless'' systems, the relaxation timescale associated to two-body encounters is significantly longer than
a Hubble time. This allows simulators to both ``soften'' the gravitational potential and use particles with masses much larger than that of individual stars.
However, the secular relaxation of smaller ``collisional'' star clusters, such as globular clusters (GCs) and galactic nuclei,  are \emph{primarily} driven by  the cumulative effect of many uncorrelated encounters along its particles' orbit \cite[e.g.,][]{Binney2008}.  The closest of these encounters may in particular lead to the formation of binaries, or even to physical collision, producing many unique stellar systems such as millisecond pulsars
\cite[e.g.,][]{Rappaport1989}, cataclysmic variables \cite[e.g.,][]{Grindlay1995}, blue stragglers \cite[e.g.,][]{Leonard1989}, and binary black hole (BH) mergers \cite[e.g.,][]{Sigurdsson1993}.
Two-body encounters are also responsible for driving stars into the center of the cluster \cite[where they may be tidally disrupted by a massive black hole, e.g.,][]{Frank1976} and out of the Lagrange points \cite[where they may escape the cluster and form stellar streams, e.g.,][]{Kupper2010}.

Because of the need to accurately model both close and distant two-body encounters, the tools employed for collisional relaxation are often significantly slower than their collisionless counterparts.
Perhaps the most natural approach to the collisional $N$-body problem is the direct summation of gravitational forces, and as such it was the first to be used \citep{Hoerner1960}. In the direct summation approach, the total force of every star upon every other star are added numerically at every timestep, providing the acceleration of all stars of the cluster.
This approach, originally limited by the computational power of the time -- e.g., 25 particles in \citet{Hoerner1960} -- has greatly benefited by the fast evolution of numerical methods and computer hardware -- reaching already 500 particles by 1973 \citep[see, e.g.,][for a review]{Aarseth1973,Aarseth1974,Heggie2003} despite its $\mathcal{O}(N^2)$ complexity. The emergence of more efficient codes executed on better hardware improved $N$-body studies. Examples of codes intensively used include the much used \texttt{NBODY} series \citep{Aarseth1985,Giersz1994a,Giersz1994b,Makino1996,Spurzem1996,Aarseth2003,Baumgardt2004,Aarseth2012,Breen2017,Pavlik2018}, along with \texttt{PhiGRAPE} \citep{Harfst2008}, \texttt{ph4} \citep{McMillan2012}, \texttt{HiGPUs} \citep{CapuzzoDolcetta2013} and \texttt{frost} \citep{Rantala2021}. Complementary approaches focused on efficiency \citep{Quinn1997,Makino2002,Wang2015, Dehnen2017,Hernandez2019} allow us to follow the evolution of a cluster of $10^6$ stars at present. Nonetheless, the large number of particles, and the close encounters which require special handling to be resolved properly, make the computational cost a lingering concern. Not only does this set practical limitations on $N$, but it  also limits our ability to perform statistical averages over many cluster realizations.

Alternative approaches have been devised to address these problems. By assuming that the evolution of stellar clusters is dominated by collisional dynamics, solutions such as Fokker--Planck methods \citep{Chandrasekhar1941,CohnKulsrud1978,Cohn1979,Takahashi1995,Drukier1999,Vasiliev2017} and Monte Carlo methods \citep{Spitzer1969,Henon1971,Shapiro1978,Hypki2013,Joshi2000,Vasiliev2015,Rodriguez2022} were developed. Both methods rely on the assumption that the secular evolution of a stellar cluster can be modeled as a succession of uncorrelated, pairwise encounters, whose cumulative effects induce a slow diffusion of each particle's mean field orbit. As a result, the mean field distribution function slowly diffuses in orbital space, and is described by a Fokker--Planck equation \citep[see, e.g.,][for a review]{Chavanis2013}. 
 
 The Fokker--Planck method aims to solve this equation directly.
Monte Carlo approaches, however, aim to study this slow evolution using a statistical approach.  These methods typically fall into one of two categories.  In an 
\emph{orbit following} (or Princeton-style) Monte Carlo, the particle trajectories are integrated in a fixed background 
potential \citep{Spitzer1971,Spitzer1972}.  As the particles advance, 
perturbations -- consistent with the diffusion coefficients of the Fokker--Planck approximation\footnote{Typically, these  coefficients are calculated assuming an 
isotropic, Maxwellian background distribution function \citep[see, e.g., appendix L of][]{Binney2008}} -- are applied to the velocities.  The most recent version of these Princeton method codes 
\citep[\texttt{Raga}, see][]{Vasiliev2015}, update the background potential as the cluster evolves using an adaptive basis expansion method, and has been used for studies of binary black hole hardening in triaxial stellar clusters. 
However, because the diffusion experienced by each particle is uncorrelated to the diffusion of every other particle, the Princeton-style Monte Carlo does not 
intrinsically conserve energy or angular momentum in its two-body dynamics, nor does it allow for the handling of close encounters responsible for the unique 
stellar and binary dynamics in GCs and galactic nuclei.

Instead, collisions and binary dynamics are more often studied by \emph{orbit-averaged} (or H\'enon-style) Monte Carlo methods. 
In \cite{Henon1971}, the author assumes the star cluster to be spherically symmetric, and calculates the gravitational potential assuming each star represents a 
spherical shell of mass at a given radius from the cluster center.  The particle positions are then updated by randomly sampling each particle's mean field orbit.  H\'enon's method assumes that the cumulative 
effects of pairwise encounters can be modeled by a single ``effective encounter'' with a \textit{close neighbor} (taken to be the particle associated 
with the 
neighboring radial shell).   This approach has been intensively explored, with the two codes most commonly in use -- \texttt{CMC} \citep{Joshi2000,Rodriguez2022} and  \texttt{MOCCA} \citep{Giersz1998,Hypki2013}  -- proving highly effective in predicting the pre- and post-core collapse behavior of GCs. Furthermore, because 
these effective encounters are performed between pairs of particles, the method   conserves energy during deflections.  Its structure also makes it possible to include additional 
physics for close encounters between neighboring particles, such as binary encounters, physical collisions, binary production, and so on. 
Nonetheless, the assumption of spherical symmetry heavily restricts the class of models which can be studied with accuracy with these methods.

Futhermore, data obtained by the HST \citep{Bellini2017} and {\it Gaia} space telescope \citep{Bianchini2018,Sollima2019} show that several popular approximations for GC modeling -- including isotropy and spherical symmetry -- that were adequate in the past \citep[see, e.g.,][]{McLaughlin2005}, are no longer satisfactory. These surveys elucidated the internal kinematics of a large sample of Milky Way GCs \citep{Bianchini2013,Fabricius2014,Watkins2015,Ferraro2018,Kamann2018, massari2024}, in addition to a quantification of velocity anisotropy in the full phase space \citep{Jindal2019}. Consequently, it is necessary to develop new tools to study the impact of anisotropy and rotation \citep[see, e.g.][]{Longaretti1997,Kim2008,Hong2013,Tep2024} but also of non-spherical objects \citep[e.g., tidal debris, see][]{Renaud2011}. The gravo-gyro catastrophe that may occur in rotating clusters \citep{Inagaki1978,Hachisu1979,Hachisu1982,Akiyama1989,Einsel1999,Ernst2007,Fiestas2010,Kamlah2022,Tep2024} warrants further exploration.

In this paper, we describe a new code, \krios\, which combines the generic 3D modeling capabilities of the Princeton method with the energy conserving and 
neighbor interaction features of H\'enon's method.
In Section~\ref{sec:theory}, we detail the theory of relaxation of globular clusters used for the implementation of the \krios\ code. Following the work of \citet{Weinberg1996,Weinberg1999,Vasiliev2015,EXP2022}, we use the self-consistent field method (SCF) to bypass the need for spherical 
symmetry and treat the relaxation of a generic cluster as the sum of its collisionless evolution and a succession of two-body deflections {\`a} la H\'enon. In Section~\ref{sec:implementation}, we detail the main components of our implementation: (i) the selection of 
particle pairs necessary for 2-body relaxation; (ii) the integration scheme used to perform collisionless relaxation; (iii) the determination of the optimal basis parameters that minimize the size of the SCF basis expansion. Finally, we reproduce in Section~\ref{sec:validation} the core collapse of an isotropic, a tangentially anisotropic and a rotating Plummer spheres and the radial orbit instability (ROI) of a radially anisotropic Plummer sphere, and compare our results against alternative methods such as direct $N$-body and a strictly collisionless code \citep[\texttt{EXP},][]{EXP2022}. We also discuss the timing of our \krios\ code.  Section~\ref{sec:conclusions} is reserved for a final discussion of our conclusions and of future works.

\section{Theoretical tools for relaxation}
\label{sec:theory}

Traditionally, H\'enon's method assumes the cluster to be spherically symmetric. This reduces the dimensionality of the problem by treating particles as triplets $(r,\vrr,\vrt)$, where $r$ is the distance of the particle to the center of the cluster, and $\vrr$ and $\vrt$ are respectively the radial and tangential velocities.  In each timestep, the particles are  sorted by increasing radii from the cluster center. This makes both the identification of neighboring particles (in a radial pairing scheme, see Section~\ref{subsec:particle_pairing}) and the calculation of the gravitational potential straightforward. To adapt H\'enon's method to arbitrary cluster geometries, we need an efficient way to determine both the \textit{local neighbors} of each particle in 3D space and a new method to calculate the gravitational potential of the cluster. We shall address each of these problems in this section.

 \subsection{The SCF method}
\label{subsec:scf_method}

 Let us begin by computing the gravitational potential and stellar mass density of the cluster. In the case of a spherical cluster, the naive quadratic complexity of the potential calculation may be reduced  to linear complexity by considering $N$ radial shells instead \citep[see, e.g.][]{Henon1971}. However, it is clear that such a method relies heavily on the assumption of spherical symmetry and, as such, cannot be applied to the case of a generic cluster. 

 To remedy this issue, we shall consider the case of a cluster with $N \gg 1$ particles and approximate its potential with a smooth mean field. Then, we shall estimate the mean field potential by expanding it over a suitable biorthogonal basis \citep{Kalnajs1971}.
Let us consider the potential-density basis elements $(\psi^{(p)},\rho^{(p)})$, where $p$ is a multi-index\footnote{In this paper, we shall use $p=(n,\ell,m)$, where $n$ is a radial number, $\ell$ a harmonic number and $m$ an azimuthal number.}. We suppose that these basis elements are biorthogonal, which means that they obey the relations
\begin{subequations}
\begin{align}
\psi^{(p)}(\br) &= \int \rd \br' \,U(|\br-\br'|) \rho^{(p)}(\br'), \label{eq:poisson_eq}\\
-\delta_{pq} &= \int \rd \br \,\psi^{(p)*}(\br) \rho^{(q)}(\br), \label{eq:bi-orthogonality}
\end{align}
\end{subequations}
where $U(r)=-G/r$ in the Newtonian interaction kernel. Equation~\eqref{eq:poisson_eq} is equivalent to the Poisson equation $\nabla^2 \psi^{(p)}=4\pi G \rho^{(p)}$, while Equation~\eqref{eq:bi-orthogonality} is the biorthogonality relation. It follows that any mean field potential and mass  density pair\footnote{We consider a cluster to be composed entirely of bound particles only, i.e. with  energies $E<0$. As such, any related computation implicitly involves the bound particles only, unless stated otherwise.\label{fn:bound_cluster}} can be written as 
\begin{subequations}
\label{eq:SCF_expansion}
\begin{align}
\label{eq:cb_potential} \psi(\br)&= \sum_p a_p\, \psi^{(p)}(\br),\\
\rho(\br)&= \sum_p a_p\, \rho^{(p)}(\br).
\end{align}
\end{subequations}
Many such basis sets have been constructed for different use cases over the years, including the study of infinitely thin discs \citep{Kalnajs1971,CB72,Kalnajs1976,Aoki1978,Aoki1979,Qian1993}, of spherically symmetric clusters \citep{CB73,Saha1991,HO1992,Zhao1996,Brown1998,Rahmati2009,Lilley2018,Lilley2023}, of flattened systems \citep{Robijn1996}  or of nearly arbitrary distributions \citep{Weinberg1999,EXP2022}.

\begin{figure} 
    \centering
    \hspace{-5mm}
    \includegraphics[width=1.0\columnwidth]{ 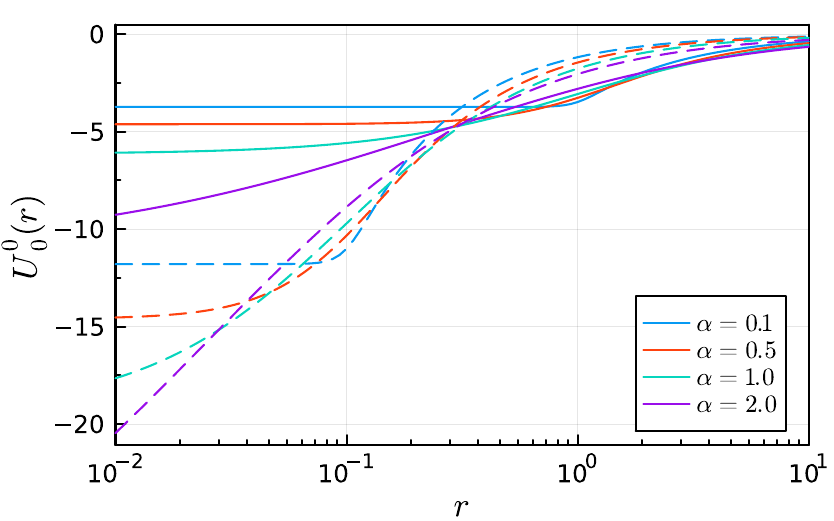}
   \caption{Behavior of the radial basis function, $U_n^{\ell}(r)$, of the lowest Zhao potential basis elements $n=\ell=0$, defined in Equation~\eqref{eq:Unl_Zhao}. We display four $\alpha$ values (shown in blue, red, green and purple by increasing order) and two $b$ values (0.1 in dashed lines, 1.0 in solid lines). The 
$\alpha$ parameter is a power-law index that sets the ``cuspiness'' of the potential-density profile. For example, $\alpha=1/2$ corresponds to a Plummer sphere and $\alpha=1$ to the Hernquist potential. The $b$ parameter can be qualitatively related to the scale length of the system.}
   \label{fig:zhao_basis}
 \end{figure}

In this paper, we shall use the parametrized basis functions described by \citet{Zhao1996}, henceforth referred to as \citetalias{Zhao1996}. The particularity of these basis elements is that they are indexed by two continuous parameters, denoted by $\alpha$ and $b$. This allows us to expand both cored and cuspy potentials by selecting the appropriate values for $\alpha$ and $b$ (see Figure~\ref{fig:zhao_basis}).
Physically, $\alpha$ is the parameter that controls how cuspy the potential basis functions are, with higher values corresponding to cuspier potentials, while the parameter $b$ corresponds to a characteristic length of the system. 
We refer to Appendix~\ref{app:coeffs_SCF} for an explicit expression of these basis elements, which take the form of a product of a radial part with a spherical harmonic.

\subsection{Calculation of accelerations}
\label{subsec:acc_calc}

\begin{figure} 
    \centering
    \includegraphics[width=0.95\columnwidth]{ 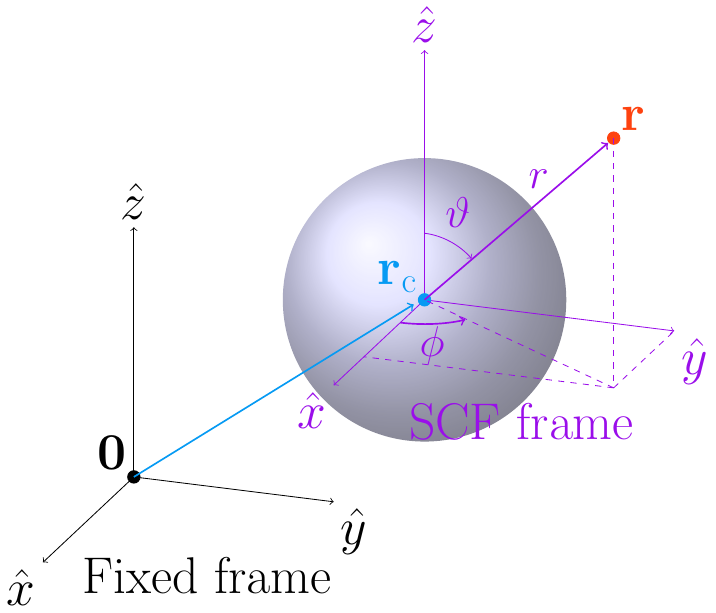}
   \caption{\krios\ uses two frames to describe the dynamics of the cluster: (i) a fixed reference frame (in black), centered at $\mathbf{0}$ and attached to a fixed set of Cartesian coordinates $(x,y,z)$; (ii) a time-dependent frame called the SCF frame (in purple), centered at the density center of the cluster $\mathbf{r}_{\rc}(t)$, attached to a translation of the same Cartesian coordinates, and to which we add a set of spherical coordinates $(r,\vartheta,\phi)$. Frame (i) is used to integrate the dynamics in an inertial frame, while Frame (ii) is adapted to the description of the cluster and is used to expand its potential-density pair using the SCF method.
   }
   \label{fig:frames}
 \end{figure}

In this paper, we shall consider two reference frames. First, we define a fixed reference frame, whose center is set to $\boldsymbol{0}$ and to which we attach the Cartesian coordinates $(x,y,z)$. Second, we define at each time $t$ the density center of the cluster $\mathbf{r}_{\rc}[t]$ \citep{Casertano1985}
\begin{align}
\label{eq:rc_ens}
    \br_{\rc}=\frac{\sum_{i=1}^N \rho(\br_i)\,\br_i }{\sum_{i=1}^N \rho(\br_i)},
\end{align}
and define the reference frame centered on $\mathbf{r}_{\rc}[t]$ such that the transformation between the two frames requires only a translation. Numerical testing showed that Equation~\eqref{eq:rc_ens} is a good proxy for the center of the cluster -- especially in the case of cuspy profiles. We also attach to this second frame -- which we shall refer to as the SCF frame -- a set of spherical coordinates $(r,\vartheta,\phi)$. We refer to Figure~\ref{fig:frames} for an illustration of these two frames.

The SCF frame is the frame that is most adapted to describe the globular cluster at a given time. As such, we shall expand the potential-density pair using the biorthogonal basis elements in that frame. 
Now, the instantaneous mass distribution of the $N$ particles is given by a sum of Dirac distributions
\begin{equation}
\label{eq:sum_dirac_dists} \rho(\br, t) = \sum_{k=1}^N m_k \, \deltaD(\br - \br_k[t]),
\end{equation}
where $\br_k[t]$ is the position of the particle $k$ at time $t$. We implicitly sum over the bound particles only, for which their total energy $E$ is non-positive (see footnote~\ref{fn:bound_cluster}). 
Since Equation~\ref{eq:sum_dirac_dists} converges to the mean field density in the large $N$-limit, the SCF expansion coefficients can be estimated by (see Appendix~\ref{app:coeffs_SCF}) 
\begin{align}
\label{eq:ap_from_N}
a_{p}(t) &= -\sum_{k=1}^N m_{k}\psi^{(p)}(\br_{k}[t])^{*}.
\end{align}
In order to speed up the potential computation, the user is able to filter out extraneous basis elements with a Boolean flag in the configuration file. Given a set of basis elements (i.e., a given value of $n_{\max}$ and $\ell_{\max}$), we sort each of them by their associated $|a_{p}|^{2}$ value and retain the most powerful coefficients such that 
\begin{align}
\label{eq:sampling_tolerance} 1 - \varepsilon \leq \frac{\sum_{{\rm retained} \, p} |a_{p}|^{2} }{\sum_{{\rm all} \, p} |a_{p}|^{2}} \leq 1,
\end{align}
where $\varepsilon$ is the required tolerance threshold. In several use cases, retaining only the zeroth-order basis element satisfies this criterion, thanks to our choice of optimized basis functions.

Finally, the acceleration in spherical coordinates (in the SCF frame) may be expressed at any location  \citep[see, e.g.,][]{Lowing2011} as
\begin{subequations}
\label{eq:dot_vsph}
\begin{align}
\dot{v}_{r}(r,\vartheta,\phi) &= -\sum_{p} a_{p} \, Y_{\ell}^{m}(\vartheta,\phi) {\rd U_{n}^{\ell}(r) \over \rd r}, \\
\dot{v}_{\vartheta}(r,\vartheta,\phi) &= -\sum_{p} a_{p} \, {\p Y_{\ell}^{m}(\vartheta,\phi) \over \p \vartheta} \frac{U_{n}^{\ell}(r)}{r}, \\
\dot{v}_{\phi}(r,\vartheta,\phi) &= - \sum_{p} a_{p} \, ({\rm i} m) \frac{Y_{\ell}^{m}(\vartheta,\phi) U_{n}^{\ell}(r)}{r\sin\vartheta}.
\end{align}
\end{subequations}
where $(r,\vartheta,\phi)$ are the spherical coordinates of the SCF frame and ${p\!=\!(\ell,m,n)}$. 
Then, we may convert these components to the Cartesian coordinates affixed to the SCF frame using the following rotation matrix
\begin{align}
\label{eq:dot_vcar}
\begin{bmatrix}
\dot{v}_{x}\\\dot{v}_{y}\\\dot{v}_{z}
\end{bmatrix} &= 
\begin{bmatrix}
    \sin\vartheta\cos\phi & \cos\vartheta\cos\phi & -\sin\phi \\ \sin\vartheta\sin\phi & \cos\vartheta\sin\phi & \cos\phi \\ \cos\vartheta & -\sin\vartheta & 0
\end{bmatrix}
\begin{bmatrix}
\dot{v}_{r}\\\dot{v}_{\vartheta}\\\dot{v}_{\phi}
\end{bmatrix}.
\end{align}
Since the two frames are described by the same Cartesian coordinates, this transformation also yields the acceleration components in the Cartesian coordinates in the fixed reference frame, where no pseudo-forces occur due to its inertial nature. 

\subsection{Two-body scattering \`a la H\'enon}
\label{subsec:twobody}

\begin{figure} 
    \centering
   \includegraphics[width=0.45\textwidth]{ 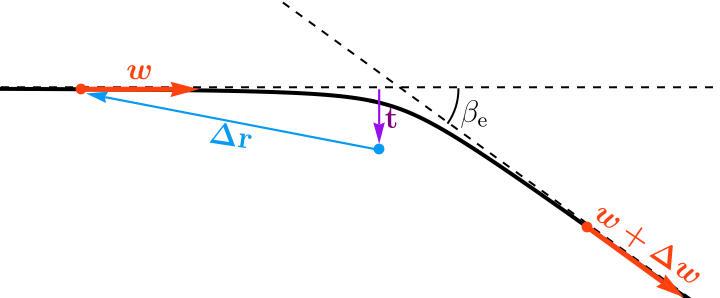}
   \caption{
   Illustration of the 2-body gravitational scattering in the plane of deflection for the particles $i$ and $i+1$, which are selected by a predetermined pairing scheme (see Section~\ref{subsec:particle_pairing}). We defined the relative position $\Delta \br=\br_i-\br_{i+1}$ and the relative velocity $\bw=\bbv_i-\bbv_{i+1}$ of the two interacting particles. These two particles are deflected by an angle $\beta_e$ and follow hyperbolic trajectories during their interaction. This results in a modification of their velocities which exactly conserves energy, but only approximately conserves angular momentum.
      }
   \label{fig:hyperbolic_2body}
 \end{figure}

We implement two-body interactions \`a la H\'enon by computing the effective scattering angle between neighboring particles and inferring the change in the velocities due to this interaction. If we denote by $i$ the index of the first particle of the interacting pair and $i+1$ the index of the second particle, then the effective scattering angle is given by\footnote{We restrict the deflection angle $\beta_{e}$ to a maximum of $\pi/2$, that is, we let $\beta_e=\pi/2$ whenever the right hand-side of equation~\eqref{eq:sin_beta_2} is greater than $1/\sqrt{2}$. We find in practice that this case occurs quite rarely.} \citep[see, e.g.,][]{Rodriguez2022} \begin{align}
\label{eq:sin_beta_2}
\sin \frac{\beta_e}{2} = \frac{\pi}{4} \sqrt{\frac{\Delta t}{T_{\mathrm{rel},i}}},
\end{align}
where $\Delta t$ is the timestep and we defined the two-body interaction relaxation timescale 
\begin{align}
\label{eq:ith_particle_dt_relax} 
T_{\mathrm{rel},i} &= \frac{\pi}{32} \frac{ w_i^3}{G^2 n(\br_i) (m_i+m_{i+1})^2 \ln \Lambda} .
\end{align}
The calculation involves the location of the pair, $\br_i$\footnote{We take this location to be that of the first particle, with index $i$ and mass $m_i$. The paired particle is given the index $i+1$ and a mass $m_{i+1}$.}, the local number density, $n(\br)$,  the Coulomb logarithm, $\ln \Lambda=\ln (\gamma N)$, and the relative velocity, $w=|\bw|$, where $\bw=  \bbv_{i}-\bbv_{i+1}$ is the relative velocity and $\gamma$ is a constant which depends on the geometry of the cluster. We set $\gamma=0.11$ for spherically symmetric cluster with equal-mass particles \citep[see, e.g.,][]{Giersz1994a,Giersz1994b}.

Thus, we need to select the interacting pair of particles $(i,i+1)$, to compute the local number density, $n(\br_i)$, and to calculate the system timestep, $\Delta t_{\rm sys}$. We discuss in Section~\ref{subsec:particle_pairing} the selection of the interacting pairs. Then, the local number density is calculated by finding the $k$-nearest neighbors ($k$-NNs) of $i$ using the method detailed in \citet{Casertano1985}, and estimating the local density as $n(\br_i)=(k-1)/V$, where $V$ is the volume of the sphere whose radius is the distance between particle $i$ and its $k^{\rm th}$ nearest neighbor. 
Finally, the selection of timestep $\Delta t$ of the integration is discussed in Section~\ref{subsec:integrator}.

Now, the two-body deflections change the orientation of $\bw$ within the plane of relative motion by the deflection angle $\beta_e$. Because \krios\ has access to the full 6D phase space, there is no need to randomly determine the orientation of the velocity vector $\bbv$, nor that of the plane of relative motion as required in the original H{\'e}non method \citep[see, e.g.,][]{Rodriguez2022}. To that end, we define the relative position vector $\Delta \br = \br_{i}-\br_{i+1}$. 
Then, we define the vector $\bt$, orthogonal to the relative velocity in the frame of relative motion, by setting
\begin{align}
\bt= (\Delta \br \cdot \hw)  \hw -\Delta \br ,
\end{align}
where $\hw=\bw/w$.  By construction, we have $\bt \cdot \bw = 0$ and $\bt \cdot (-\Delta \br) > 0$, therefore the effective particle is deflected in the direction of $\bt$ (see Fig.~\ref{fig:hyperbolic_2body}). 
Then, we can write the final relative velocity of the pair after the hyperbolic encounter as
\begin{align}
\bw^{\mathrm{new}} = \bw \cos \beta_e + w \sin \beta_e  \hrt,
\end{align}
where $\hrt = \bt/|\bt|$. Thus, the new velocities of the two particles (in the cluster frame) are given by
\begin{subequations}
\begin{align}
\bbv_i^{\mathrm{new}} &= \bbv_i + \bigg(\frac{m_{i+1}}{m_i+m_{i+1}}\bigg) (\bw^{\mathrm{new}}-\bw),\\
\bbv_{i+1}^{\mathrm{new}} &= \bbv_{i+1} - \bigg(\frac{m_{i}}{m_i+m_{i+1}}\bigg) (\bw^{\mathrm{new}}-\bw).
\end{align}
\end{subequations}
Although the scattering angle is calculated from the average over many encounters, it is applied as a single ``effective'' two-body encounter between neighboring particles.  As such, it can be shown that, by construction, this process conserves the total energy of the pair and thus the total energy of the cluster, but only conserves their total angular momentum approximately due to the cluster's finite extent. 

\section{Implementation}
\label{sec:implementation}

\begin{figure*} 
    \centering
    \includegraphics[width=0.95\textwidth]{ 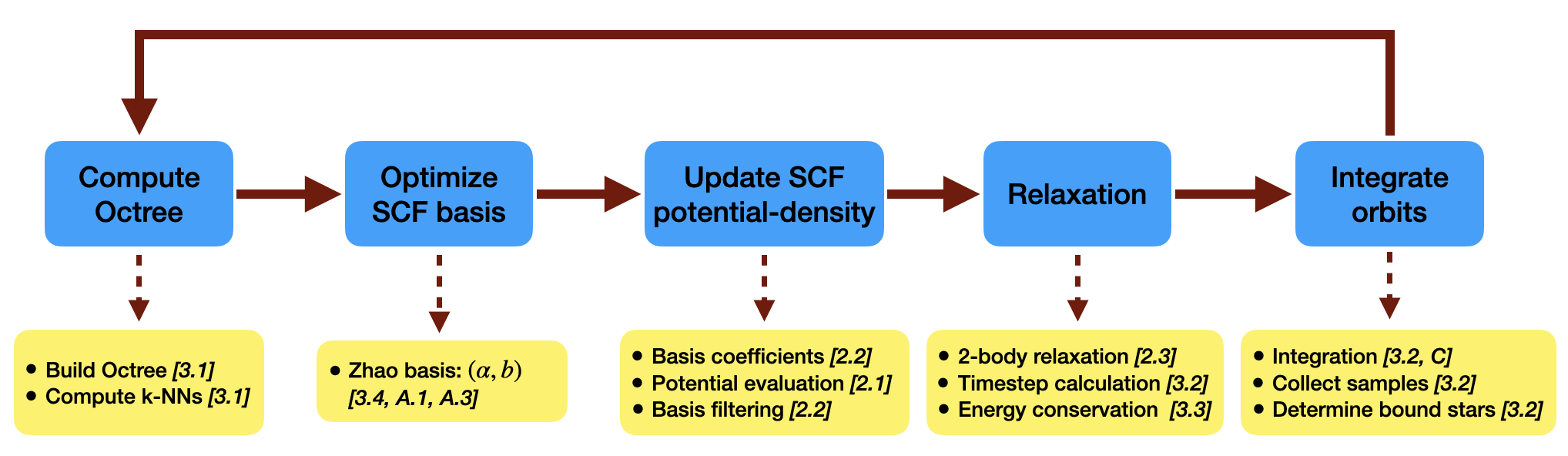}
   \caption{Schematic flowchart of the \krios\ code during one timestep. The  sections describing each entries are given within brackets.  }
   \label{fig:schema_code}
 \end{figure*}

  \subsection{Particle pairing}
\label{subsec:particle_pairing}

Although the SCF method treats the particles as $N$ independent integrations, H\'enon's scheme for two-body relaxation (see Section~\ref{subsec:twobody}) requires an up-to-date list of neighbors for each particle.  This is needed for the pairing of particles (for the effective encounters), as well as for the calculation of locally-averaged quantities (e.g., the number density in Equation~\ref{eq:ith_particle_dt_relax}).  In orbit-averaged implementations of H\'enon's algorithm, both are accomplished during the radial sorting of particles. The particles are paired in order of increasing radius --- particles 1 and 2 are paired, as are particles 3 and 4, and so on.  The averages are computed similarly using $k$-NNs in radial space.  In \texttt{cmc}, for instance, the relative velocity dispersion and the number density at particle $i$ are calculated over 40 neighbors in radius, from particle $i-20$ to $i+20$.

However, \krios\ tracks the full 6D phase space of every particle, which offers additional flexibility in both the pairing of particles and the calculation of the local averages.  To compute these averages, we calculate the $k$-NNs around each particle in the 3D space.  This search is greatly accelerated by an octree data structure, which we construct based on the position of the particles in every relaxation timestep.  The octree is a three-dimensional tree data structure that recursively partitions space into eight smaller child trees. Each child tree is recursively subdivided until at most one particle remains in each leaf. Unlike most tree-based gravity solvers \cite[e.g.,][]{Barnes1986} in which the tree is updated as each particle moves along its orbits, the octree in \krios\ is only updated  on a relaxation timescale, meaning that the particles have undergone multiple orbits since their last update (effectively moving every particle in the cluster to a new branch of the tree).  As such, it is more efficient to construct the tree from scratch every timestep, than to delete and re-insert particles into the pre-existing octree structure.

With the octree constructed, we recursively search around each particle for its $k$-NNs in space. To search for neighbors, we first calculate around each particle an initial guess of the minimum radius of a ball ($R_{\mathrm{min},k\mathrm{NN}}$) that would enclose $k$ neighbors.  Since this radius should be proportional to $\rho^{-1/3}$ at that point, we start our search using the octree to identify all particles that lie within a sphere with radius $R_{\mathrm{min},k\mathrm{NN}}= A\rho^{-1/3}$ via a recursive, top-down search of the octree.  The constant $A$ is the ratio between the true density calculated using the $k$-NN search and the SCF density averaged over all particles in the cluster.  It is updated every timestep and used to begin the search in the next timestep (as the cluster density profiles do not change substantially between timesteps).  If our original guess for $R_{\rm{min}~kNN}$ contains less than $k$ particles, we increase the search radius incrementally until we have identified a region containing $k$ particles. Testing indicates that our SCF-optimized $k$-NN search is comparable in speed to publicly available parallelized $k$-NN algorithms \cite[e.g.,][which uses a KDTree to accelerate the search for $k$-NNs]{Dalitz2009}.

With the local averages computed, we must still decide on a pairing scheme.  Like in H\'enon's original algorithm, these particles must be sufficiently close so that the neighbors are a fair draw from the distribution function at that point (essentially sampling the distribution function in true Monte Carlo fashion).  \krios\  offers two different methods for pairing particles:

\begin{itemize}
\item \textbf{Radial Pairing} -- where the particles are sorted by increasing radii from the center of the cluster (taken to be the center of the SCF frame.  
\item \textbf{Density Pairing} -- where the particles are linked in terms of decreasing density.  In this method, we start with the particle with the largest local density in the cluster, and pair it with whichever of its $k$ neighbors has the next highest density.  This chain is linked until every particle in the cluster has been incorporated. \end{itemize}

On the one hand, although the radial pairing scheme is closer to H\'enon's original method, applying it in this context will lead to the pairing of stars in radically different spatial locations, and therefore with different velocity distribution functions \textit{a priori}. Testing showed that this dramatically broke the conservation of the angular momentum vector (see Appendix~\ref{app:iom_radial_v_density}), and in turn made impossible to study clusters in rotation.

On the other hand, numerical experimentation showed that the density pairing method produced better results for isolated, core-collapsing systems. As such, we will be using the density pairing method in this paper.

\subsection{Timestep Mechanics and Orbit Integration}
\label{subsec:integrator}

We update the SCF at each system timestep. This involves flagging ${E\geq0}$ particles as unbound to the cluster such that their contribution to the gravitational field is ignored, re-computing the SCF center of the bound particles \citep[i.e., center of density,][]{Casertano1985}, and re-computing the \citetalias{Zhao1996} parameters ${(\alpha,b)}$, as well as the associated weights $\{a_{p}\}$ (see Appendix~\ref{app:scf} for more details). The system timestep depends on the parameters of the \krios\ simulation:
\begin{itemize}
    \item In collisionless runs, we set ${\Delta t_{\rm sys} = 0.05 \, t_{\rm{end}}}$.  The user may enforce a shorter timestep ${\Delta t_{\rm sys} = \Delta t_{\rm user}}$ in the configuration file.
    \item If H\'{e}non-style two-body relaxation is considered, we set ${\Delta t_{\rm sys} = \min(t_{\rm rh},\hat{T}_{\mathrm{rel}} )}$, where $t_{\rm rh}$ is the half-mass relaxation time (see Equation~\ref{eq:trh}) and $\hat{T}_{\mathrm{rel}}$ is the two-body relaxation timescale (see Equation~\ref{eq:Trel_avg}). \end{itemize}
In order to compute the two-body relaxation timescale, we first estimate the local average relaxation timescales, $\hat{T}_{\mathrm{rel},i}$, and then take their minimum \citep{Rodriguez2022}. These local timescales are the averaged versions of the relaxation timescales given in Equation~\eqref{eq:ith_particle_dt_relax}
\begin{subequations}
\label{eq:Trel_avg}
\begin{align}
\hat{T}_{\mathrm{rel,i}} &= \frac{\pi}{32} \frac{ \langle w_i\rangle^3}{4 G^2 n(\br_i) \langle m_i^2 \rangle \ln \Lambda},\\
\hat{T}_{\mathrm{rel}} &= \min_i \,\hat{T}_{\mathrm{rel,i}} .
\end{align}
\end{subequations}
They require the calculation of two averages: (i) the mean relative velocity amplitude, $\langle w_i\rangle$, which is estimated by computing the average value of $|\bbv_i-\bbv_k|$ over the $k$-NNs of $i$; (ii) the mean square mass, $\langle m_i^2\rangle$, which is computed by taking the average value of $m_k^2$ over the same $k$-NNs.

During the integration step (Figure~\ref{fig:schema_code}, ``Integrate orbits"), all particles are evolved forward in time by one timestep ${t\to t+\Delta t_{\rm sys}}$ using the SCF expansion coefficients at time $t$. This permits simple parallelization \cite[e.g.,][]{Hernquist1995}, as each particle's equations of motion depend exclusively on the mean field (see Appendix~\ref{app:integrator}). Each particle then re-synchronizes at time ${t+\Delta t}$, at which point we evaluate certain stopping conditions -- e.g., if there are any remaining bound particles --  before re-evaluating the SCF. 

It is possible to reduce the noise induced by the finite number of particles in the system by replacing the density profile by a temporal average taken over the previous timestep \citep{HO1992, Vasiliev2015} via
\begin{align}
\label{eq:rho_tn_int}
    \rho(\br,t_n)\simeq\frac{1}{t_n-t_{n-1}}\int_{t_{n-1}}^{t_n} \rd t\,\rho(\br,t).
\end{align}
In practice, we uniformly collect $N_{\rm samples}$ position samples from each particle over the previous timestep, where ${N_{\rm samples} = \max\left(1, \lceil n \, (\Delta t_{\rm sys} / t_{\rm dyn}(R_{\rc}))\rceil\right)}$ with $R_{\rc}$ the core radius (see Equation~\ref{eq:Rc}) and $n$ a numerical parameter set in the  configuration file. This allows us to estimate Equation~\eqref{eq:rho_tn_int} as 
\begin{align}
\label{eq:rho_tn_sum}
    \rho(\br,t_n)&\simeq\frac{1}{N_{\rm samples}}\sum_{j=1}^{N_{\rm samples}} \rho(\br,t_j).
\end{align}
Using the instantaneous density profile, Equation~\eqref{eq:rho_tn_sum} reduces to
\begin{align}
    \rho(\br,t_n)&\simeq \sum_{k=1}^N\frac{m_k}{N_{\rm samples}}\sum_{j=1}^{N_{\rm samples}} \deltaD(\br-\br_k[t_j]).
\end{align}
This reformulation yields the SCF coefficients 
\begin{align}
\label{eq:sampling} a_{p} \simeq -\sum_{k} \frac{m_{k}}{N_{\rm samples}} \sum_{j=1}^{N_{\rm samples}}  \psi^{(p)}(\br_{k}[t_{j}])^{*},
\end{align}
where $z^*$ is the complex conjugate of $z$. These coefficients are less impacted by the system's noise.

Now, in order to evaluate Equation~\eqref{eq:sampling}, we collect intermediate locations of the bound particles by breaking up the integration step into $N_{\rm samples}$ substeps (see Section~\ref{subsec:acc_calc}). We used an embedded $8^{\rm th}$-order Runge--Kutta stepper \citep[][]{dormand1980}, leaving the usage of different integrators \citep[e.g.][]{rein2015} to future works. The identification of unbound particles and their subsequent dynamics are critical to \krios's tidal debris modeling capabilities and will be addressed in more detail in \cite{Cook2025}.

\subsection{Energy conservation and \stod\ prescription}

The total energy of the cluster, as a function of time, is
\begin{align}
E(t) &= {1\over 2}\left(\sum_{k}m_{k}v_{k}^{2}(t) + \sum_{k} m_k \tilde{\psi}(\br_k[t],t)\right),
\end{align}
where $\tilde{\psi}(\br_k)=\psi(\br_k)+\psi_{\mathrm{sg},k}$  accounts for the self-gravity of the $k^{\rm th}$ particle  (see Appendix~\ref{subsec:self_gravity}). 
To ensure adequate energy conservation, we employ a modified version of the \cite{stodolkiewicz1982} prescription to take the time variability of the SCF into account. 

Following Appendix~\ref{app:stod_prescription}, the energy change that the $k^{\rm th}$ particle experiences as a result of mean field variability can be expressed as

\begin{subequations}
\label{eq:stod_eq}
\begin{align}
\Delta E_{k} &= {m_{k} \over 2}\Big(\Delta\psi[\br(t+\Delta t)] + \Delta\psi[\br(t)]\Big), \\
\Delta \psi(\br) &\equiv \psi_{\rm current \, SCF}(\br) - \psi_{\rm prev \, SCF}(\br).
\end{align}
\end{subequations}
In most cases, we compensate the energy variation $\Delta E_{k}$ by slightly modifying the kinetic energy of the $k^{\rm th}$ particle. However, there are instances in which this prescription yields a negative kinetic energy. If so, then the particle's kinetic energy is left unchanged, and we add the energy variation to a total budget, $E_{\rm error}$. This budget is then distributed throughout the entire system by modulating each bound particle's kinetic energy as
\begin{align}
v_{i,{\rm new}}^{2} &= \left(1 - \frac{2 E_{\rm error} }{ \sum_{k}  m_{k} \, v_{k, {\rm old}}^{2}} \right) \, v_{i, {\rm old}}^{2}.
\end{align} 
In practice however, very few particles actually contribute to this budget. Decreasing the timestep $\Delta t$ also reduces the need to this energy compensation, because the approximation used to compute the energy change becomes better (see Appendix~\ref{app:stod_prescription}).

\subsection{Optimization of the Zhao parameters}
\label{subsec:opti_zhao_alpha_b}

\begin{figure} 
    \centering
    \hspace{-4mm}
    \includegraphics[width=0.49\textwidth]{ 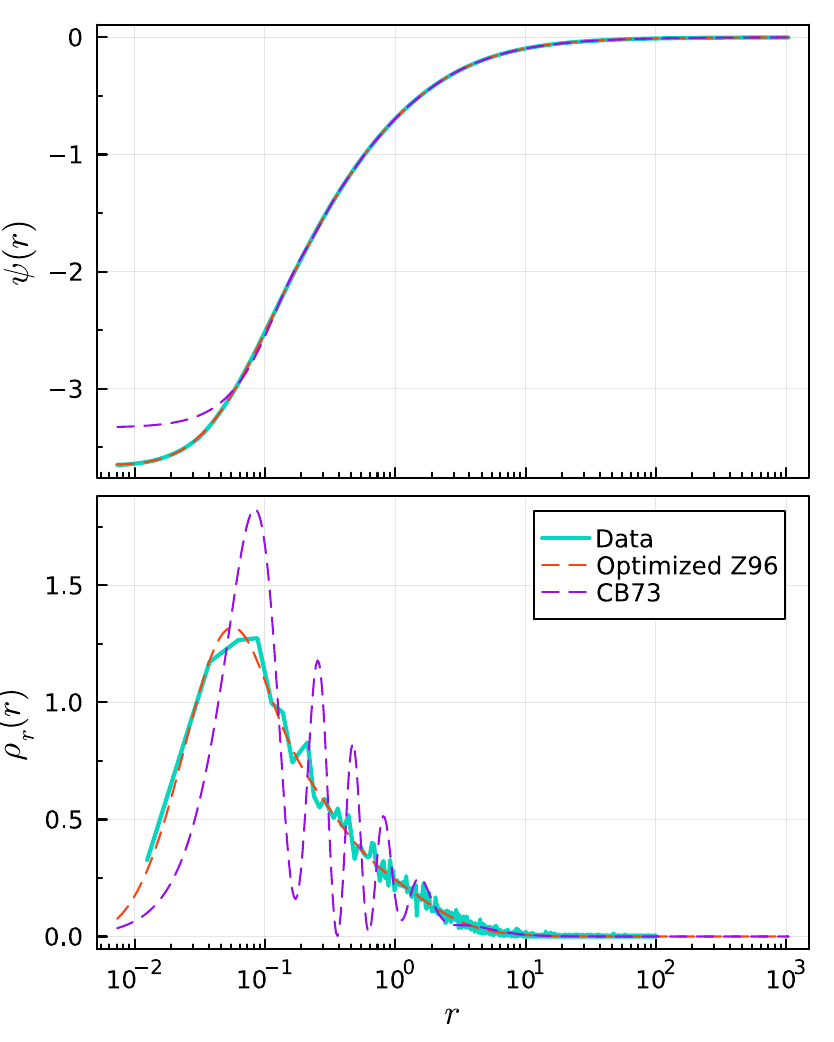}
\caption{Profile of the potential (top panel, $\psi$) and radial density (bottom panel, $\rho_r$) for a cluster nearing core collapse. The instantaneous potential and density are represented in green, while the SCF expansions are represented in red (using an optimized \citetalias{Zhao1996} family) and in purple (using the \citetalias{CB73} family). Although both radial expansions are limited by $n_{\max}=10$, it is clear that the \citetalias{CB73} expansion struggles to converge towards the instantaneous potential-density pair, whereas the optimized \citetalias{Zhao1996} expansion quickly converges towards it, illustrating both its necessity and effectiveness. }
   \label{fig:zhao_fitting}
 \end{figure}

In theory, the SCF method should allow us to expand the potential-density pair of any system, provided that we use enough basis elements. However, because the calculation of its coefficients is the primary computational cost during an integration timestep, the efficiency of the SCF method relies on using as few basis elements as possible \citep[e.g.,][]{HO1992}.  Should the basis family not be close enough to the potential-density pair of system of interest, the number of elements required for the expansion to converge may become impractical. Figure~\ref{fig:zhao_fitting} represents the potential-density pair for a Plummer cluster in the midst of core collapse and illustrate how slow the SCF method is if we use the (ill-tailored) \citetalias{CB73} basis elements. In practice, we want to minimize the number of elements needed to recover the potential precisely enough. For that purpose, we use a parameterized basis family -- the \citetalias{Zhao1996} basis functions  -- and optimize its parameters, $(\alpha,b)$, to tailor them to the instantaneous potential. 
Figure~\ref{fig:zhao_fitting} shows how an appropriate choice of parameters $\alpha$ and $b$ speeds up the convergence of the SCF expansion.
 We refer to Appendix~\ref{app:opti_Zhao} for the details of this optimization.  
 
\begin{figure} 
    \centering
    \includegraphics[width=0.49\textwidth]{ 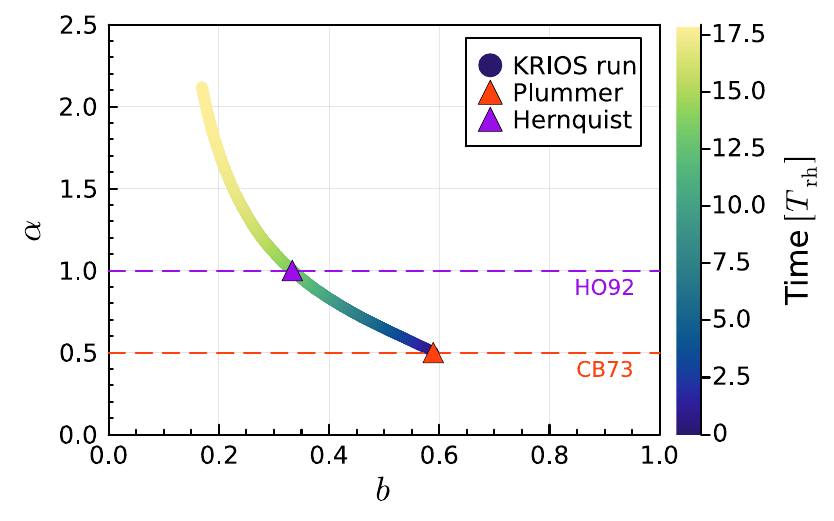}
     \caption{ Evolution of the \citetalias{Zhao1996} basis parameters, $\alpha$ and $b$, as an isotropic Plummer sphere with $N=10^6$ particles undergoes core collapse. In order for the \citetalias{Zhao1996} basis functions make it possible to accurately reproduce the mean field potential of the cluster as it becomes more and more cuspy over the course of its relaxation, higher values of $\alpha$ and lower values of $b$ are necessary. The dashed red line corresponds to the  \citetalias{CB73} family ($\alpha=1/2$), whereas the dashed  purple line corresponds to the  \citetalias{HO1992} family ($\alpha=1$). 
   The colored triangles represent the best-fit parameters corresponding to a Plummer sphere (in red) and a Hernquist sphere (in purple) in H{\'e}non units. }
   \label{fig:Zhao_parameters}
 \end{figure}

 Of course, these basis parameters  depend on the geometry of the cluster and, as such, will evolve as the cluster undergoes core collapse. 
As discussed in Section~\ref{subsec:scf_method} and shown in Figure~\ref{fig:zhao_basis}, increasing values of $\alpha$ correspond to increasingly cuspier potentials, whereas decreasing values of $b$ correspond to smaller systems. As such, we expect $\alpha$ to grow and $b$ to decrease along the core collapse process. This is exactly what we observe in  Figure~\ref{fig:Zhao_parameters}. 
We denote a slow monotonic increase of $\alpha$, which increases very quickly during the last fractions of $\trh$ of the core collapse process. Similarly, the value of $b$ decreases toward 0 during the relaxation of the cluster, with an acceleration of its decrease in the late stages of core collapse.

  \section{Results}
\label{sec:validation}

We use \krios\ to model two well-known properties of evolving, initially spherical star clusters: (i) the collisionless emergence of the radial orbit instability (and its associated triaxial mass distribution) in a radial anisotropic Plummer sphere (Section~\ref{subsec:roi_q_2}); (ii) the core collapse of an isotropic Plummer sphere (Section~\ref{subsec:CC_q_0}), a tangentially anisotropic one (Section~\ref{subsec:cc_q_-6}) and a rotating isotropic one (Section~\ref{subsec:relaxation_q_0_alpha_0.5}).  We compare our results to $N=10^4$ particle clusters evolved with \texttt{NBODY6++GPU} \citep{Wang2015} -- as described in \citet{Tep2022,Tep2024} -- though alternative comparisons will be shown for the ROI.

\begin{figure} 
    \centering  \includegraphics[width= \columnwidth]{ 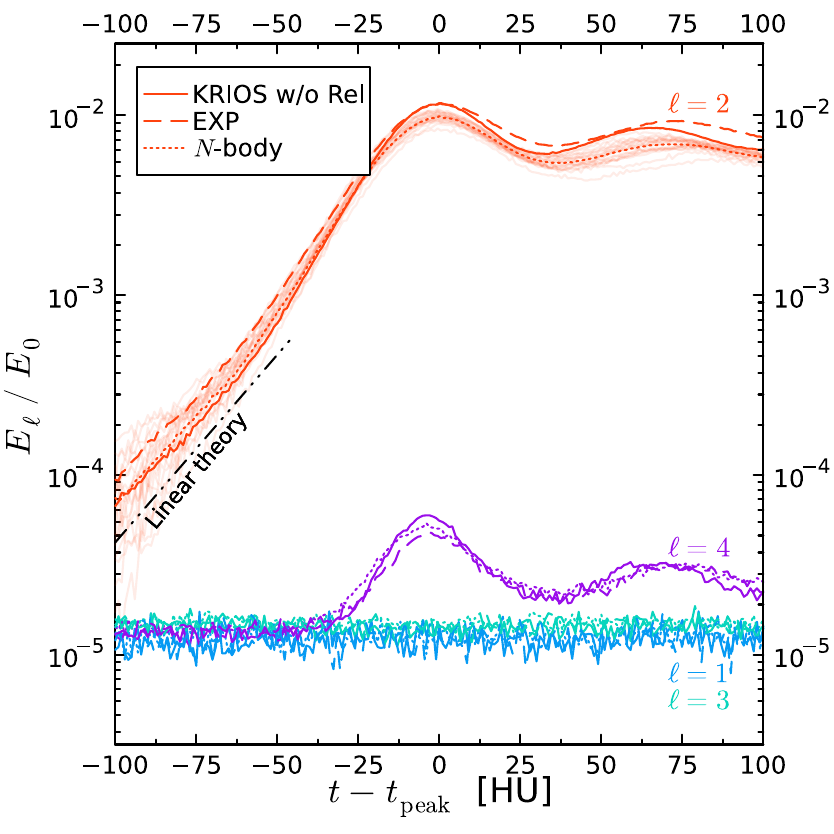}
   \caption{
   Evolution of the normalized harmonics amplitude, $E_{\ell}/E_0$, measured in $N$-body runs (dotted  lines, averaged over 19 runs), in collisionless \krios\ runs (full lines, averaged over 10 runs, with timestep $\Delta t_{\rm sys}=0.01 \, \rHU$) and (collisionless) \texttt{EXP} runs (dashed-dotted lines, averaged over 10 runs). The growth rate prediction by linear theory using the Julia package \href{https://github.com/JuliaStellarDynamics}{\texttt{JuliaStellarDynamics}} is shown as a dashed black line along the curve. All runs contain $N=10^5$ particles of equal masses.
   We centered the time axis at the peak of the $\ell=2$ amplitude for each realization before ensemble average. The ROI drives the amplification of the quadrupole $\ell=2$ and, in a lesser fashion, that of the $\ell=4$ component, until saturation occurs. 
   Two-body relaxation tends to counteract ROI, leading to a slightly lower quadrupole amplitude and a slow decrease over time similar to $N$-body measurements (as opposed to the constant saturation in collisionless runs). 
   The other harmonics are not impacted by the ROI, and therefore remain at the noise level $\sim 1/N=10^{-5}$. Finally, collisionless runs (\krios\ w/o 2-body relaxation and EXP) closely recover the ROI observed in direct $N$-body runs. This is because the 2-body relaxation timescale computes to $\hat{T}_{\mathrm{rel}}\sim 200 \, \rHU$ (at $t=0$), which is much longer than the dynamical timescale over which the ROI occurs. 
      }
   \label{fig:E_ell}
 \end{figure}

 \subsection{Radial orbit instability (ROI)}
 \label{subsec:roi_q_2}

In this section, we test \krios\ in the regime of non-spherical systems. To that end, we set our initial conditions as a $q=2$ radially anisotropic cluster \citep{Dejonghe1987} composed of $N=10^5$ equal-mass particles (see Appendix~\ref{app:plummer_cluster} for details), and will be studying their \textit{collisionless} evolution. 
It is known that this cluster is subject to the ROI \citep[see, e.g.,][]{Palmer1994,Polyachenko2015,Petersen2024} due to its overabundance of radial orbits \citep{Binney2008,Marechal2011}, and therefore quickly transitions from a spherically-symmetric state to a triaxial one. 

 Following \citet{Petersen2024}, we set the timestep of our code to be $\Delta t_{\rm sys} = 0.01 \, \rHU$ in order to resolve this instability.
We may quantify the ROI's impact by looking at the evolution of the quadrupole mode ($\ell=2$). For that purpose, we define
\begin{align}
    E_{\ell} = \sum_{n,m} |a_{\ell m n}|^2,
\end{align}
where $a_{\ell m n}$ are the coefficients of the SCF expansion given by Equation~\eqref{eq:SCF_radial_harmonics}. By construction, $E_{\ell}$ is the energy contained in the $\ell^{\rm th}$ harmonics. We show in Appendix~\ref{app:amplitude_ell} that this quantity does not depend on the radial basis family, nor on the orientation of the coordinate system.

The resolution of the ROI by different models -- \texttt{NBODY6++GPU} \citep{Wang2015}, collisionless \krios, and \texttt{EXP} \citep{EXP2022} -- is shown in Figure \ref{fig:E_ell}, and is directly compared against theoretical predictions from linear response theory (\href{https://github.com/JuliaStellarDynamics}{https://github.com/JuliaStellarDynamics}). In particular, we compare the early evolution of $E_{\ell}/E_{0}$ for the ${\ell\leq4}$ harmonics.
Each time series dataset is centered on the moment at which $E_{\ell=2}$ is maximized, and each run is then averaged with respect to that time centering. We refer to Table~\ref{table:GR_ROI} for the details of the runs.
\begin{table}[h!]
\centering
\setlength{\tabcolsep}{.3em}
 \begin{tabular}{|c|c|c|c|c|c|}
\hline
Method & Growth rate &$N$ & $N_{\rm runs}$ &  $n_{\max}$ & ${\ell}_{\max}$ \\
\hline
\hline
\krios\ w/o rel.   & 0.0229  &  $10^5$ & $10$  &   12 & 6 \\
$N$-body  & 0.0227  &  $10^5$ & $19$  &   $\emptyset$ & $\emptyset$ \\
\texttt{EXP}  & 0.0229  &   $10^5$  &  10   &    12  &  6 \\
Linear theory & 0.0240  &  $\emptyset$  &  $\emptyset$   &    30  &  2 \\
\hline
\end{tabular}
\caption{Values of the $\ell=2$ growth rate of the ROI obtained via collisionless \krios\ runs, \texttt{EXP} runs, $N$-body runs (\texttt{NBODY6++GPU}) and linear theory. Quantities are shown in H\'enon units, and agree within 5\%.
}
\label{table:GR_ROI}
\end{table}

First, we observe exponential growth between $t=-100 \,\rHU$ and $t\sim-60 \,\rHU$ for all three numerical simulations. Linear regression allows us to compute the 
slope ${\rd(\ln {E}_{\ell=2})/\rd t \equiv 2 \gamma}$, where $\gamma$ is the growth rate. Table~\ref{table:GR_ROI} shows that the numerical simulations and the theoretical prediction from linear response theory agree within 5\%. After $t\sim-60 \,\rHU$, the system enters the non-linear regime and reaches saturation at $t \sim 0\, \rHU$, before the ${\ell>0}$ components slowly decay via damped oscillations.

Overall, \krios\ appears to be consistent with the alternative codes we considered and is especially consistent with the measurements from $N$-body simulations. In fact, the close similarity between the collisionless codes (both our collisionless \krios\ runs and the \texttt{EXP} ones) and the $N$-body runs may be understood by computing the 2-body relaxation time given by Equations~\eqref{eq:ith_particle_dt_relax}. Applying this equation to the case of the $q=2$ isotropic Plummer cluster yields a 2-body relaxation timestep of order $\hat{T}_{\mathrm{rel}}\sim 200 \, \rHU$, hence a timescale much longer than the dynamical time during which the ROI occurs.

\begin{figure}[t] 
    \centering
\includegraphics[width=0.95 \columnwidth]{ 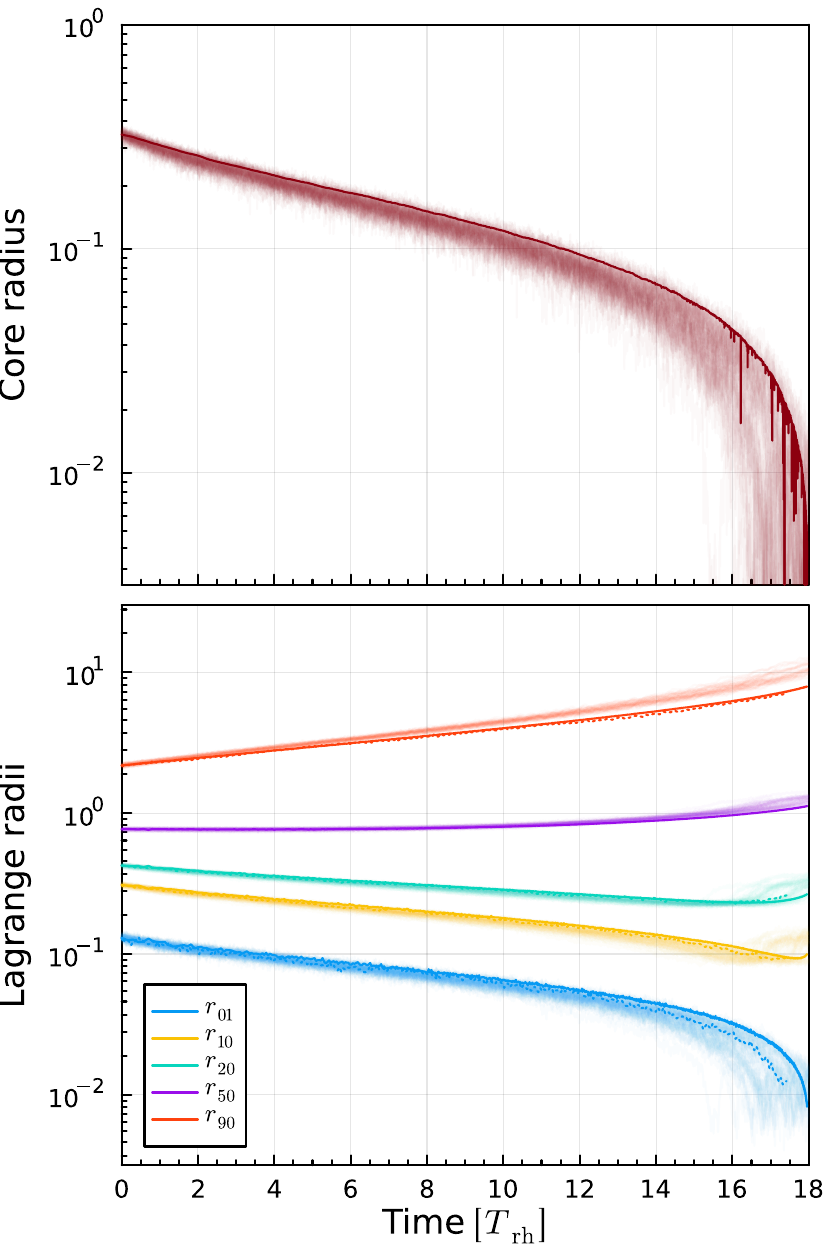}
   \caption{ Top panel: Evolution of the core radius (Equation~\ref{eq:Rc}) of an initial isotropic, non-rotating Plummer sphere (with identical masses) using $N$-body simulations (in transparent colors, $N=10^4$)  and the \krios\ code (in red, $N=10^6$ and $n_{\mathrm{neigh}}=30$, $\ell_{\max}=0$ and $n_{\max}=15$). Bottom panel: Evolution of a few Lagrange radii for the same system using $N$-body simulations (transparent lines), the \krios\ code (solid lines). The dotted lines corresponds to the \raga\ run described in figure~2 of \citet{Vasiliev2015} and rescaled to our units.  Time is expressed in units of initial half-mass relaxation time (see Equation~\ref{eq:trh}). We show a sample of 50 realizations of $N$-body simulations. The SCF approach is consistent with measurements made in $N$-body simulations over the course of the cluster's relaxation up to the deeper stages of core collapse. Using the SCF method, \krios\ is able to reach core collapse in a reasonable time ($t_{\rc\rc} \sim 18 \,\trh$)  -- a value that weakly depends on the run parameters. }
   \label{fig:core_collapse_Rc_lag}
 \end{figure}

\begin{figure} 
    \centering
    \hspace{-4mm}
\includegraphics[width=0.49\textwidth]{ 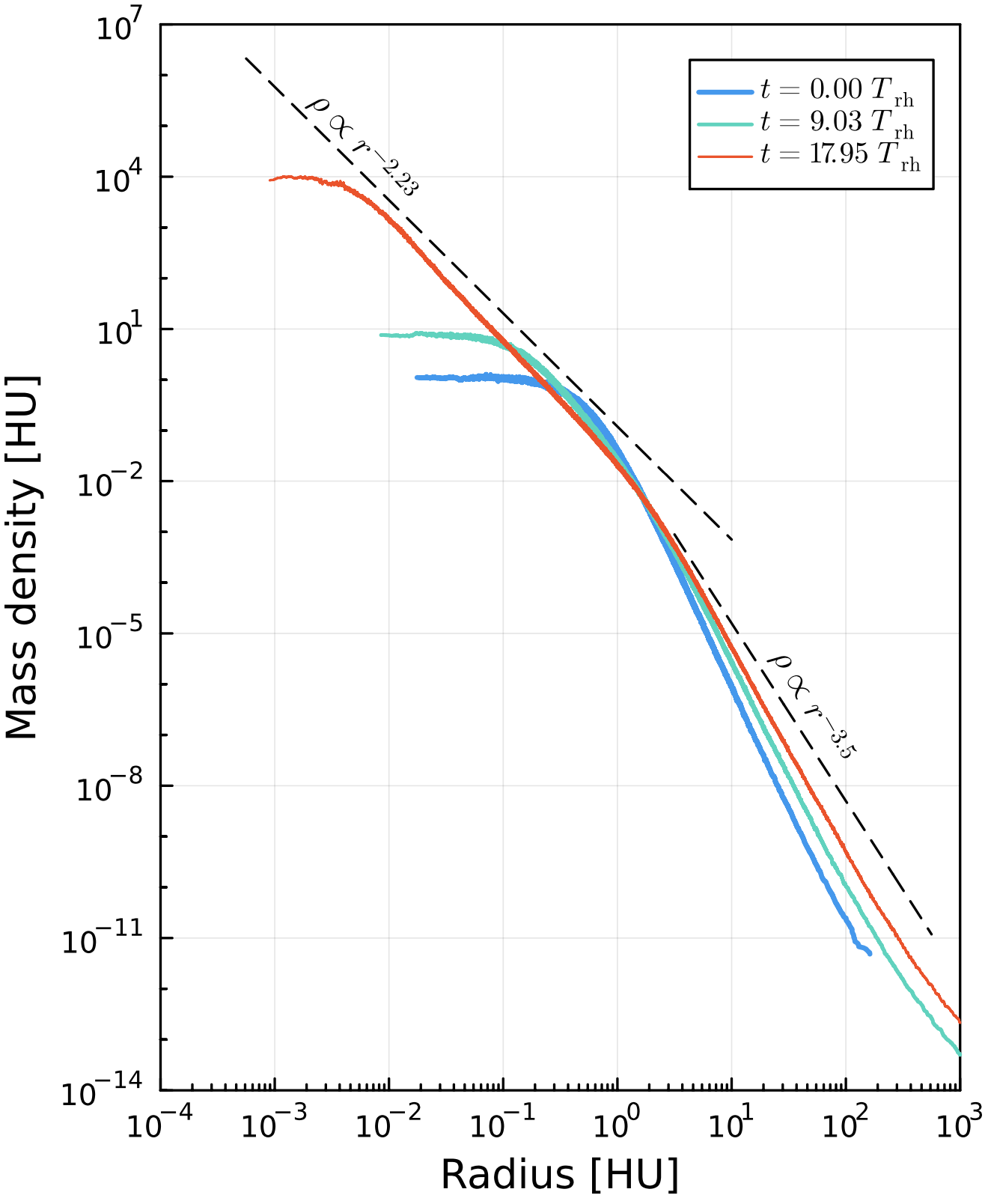}
\caption{ Evolution of the SCF three-dimensional local mass density, $\rho(\br)$, for the run shown in Figs.~\ref{fig:core_collapse_Rc_lag} and 
\ref{fig:IOM}. As we near core collapse, density tends to converge towards the $r^{-2.23}$ behavior in the central region and $r^{-3.5}$ behavior in the outer region, which is in agreement with 
\citet{Takahashi1995}. }
   \label{fig:core_collapse_density}
 \end{figure}

\begin{figure} 
    \centering
    \includegraphics[width=0.95 \columnwidth]{ 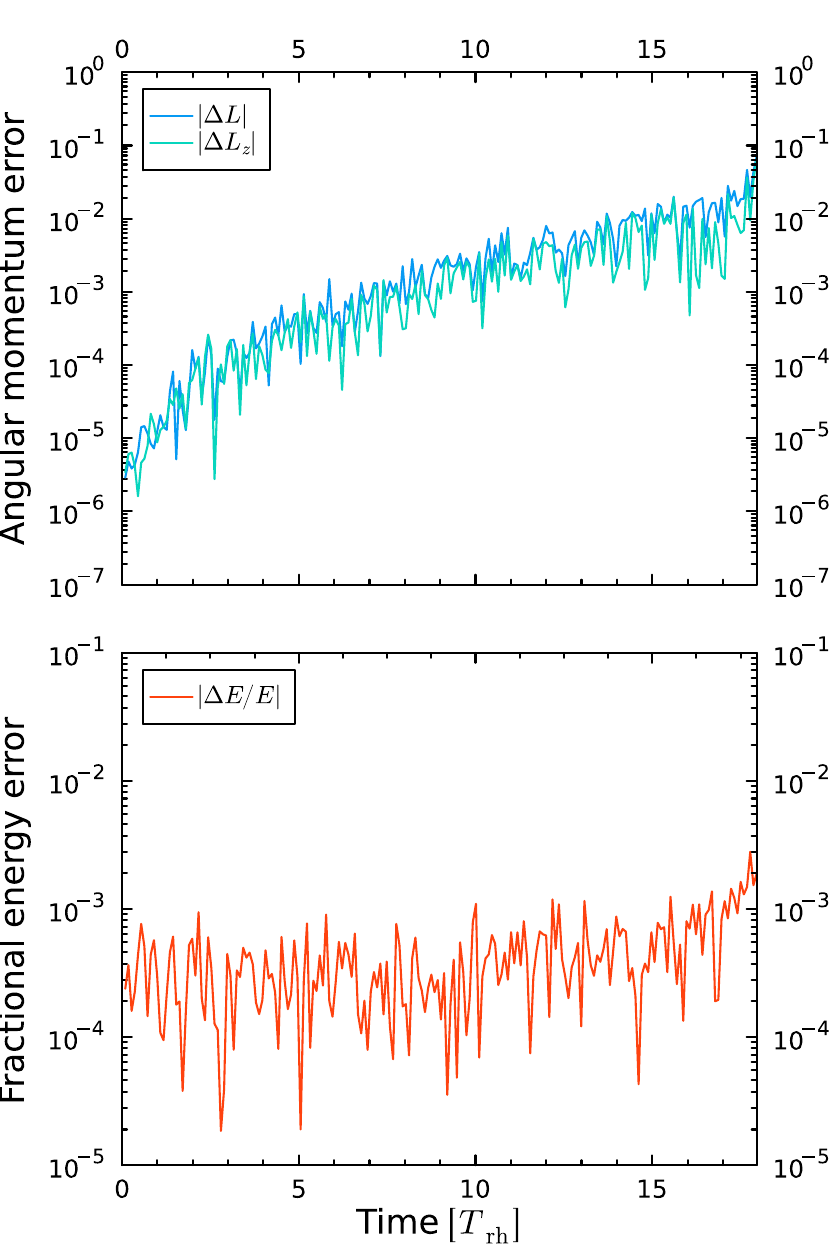}
   \caption{ Evolution of the fractional energy change (bottom panel) and of the absolute error in angular momentum (in H{\'e}non units), $|\Delta L|, |\Delta\Lz|$, during the cluster's evolution shown in Fig.~\ref{fig:core_collapse_Rc_lag}, in the SCF frame. Energy conservation is satisfactory due to the \stod\ scheme adjustment, and  stays well below $1\%$   relative error at core collapse. Similarly, angular momentum conservation remains below $|\delta \bL| \leq 0.06\,\rHU$ -- despite the \stod\ prescription and the approximate conservation in 2-body collisions (see Section~\ref{subsec:twobody}) -- provided we use the density pairing scheme (see Section~\ref{subsec:particle_pairing}).
      }
   \label{fig:IOM}
 \end{figure}

\subsection{Relaxation of an isotropic Plummer sphere}
  \label{subsec:CC_q_0}

We use a Plummer sphere (see Appendix~\ref{app:plummer_cluster}) containing  $N$ equal-mass particles as our initial conditions. The initial half-mass relaxation time, $\trh$, is given by
\begin{align}
\label{eq:trh}
    \trh &= \frac{0.138 \,N^{1/2}\, \rrh^{3/2}}{(G m)^{1/2} \ln(0.11 N)},
\end{align}
where $\rrh$ is the (initial) half-mass radius and $m$ the individual stellar mass \citep[We refer to section~14 of][ for more details]{Heggie2003}. For an $N=10^{6}$ 
cluster, we have $\trh=7\, 966$ H\'{e}non units (HU). The HU system is defined such that ${G=M=R_{\rm v}=1}$, where $R_{\rm v}$ is the virial radius. It follows that the total energy of the system is $E=-0.25$. We use 1 orbital 
sample per dynamical time, $k=30$ neighbors for the local number density estimates, set the maximum number of basis elements to $n_{\max}=15$ and $\ell_{\max}=0$, and set the threshold for basis filtering at $\varepsilon=10^{-6}$ (see Equation~\ref{eq:sampling_tolerance}). Finally, we use the density pairing scheme described in Section~\ref{subsec:particle_pairing} to determine how to apply the 2-body relaxation step detailed in Section~\ref{subsec:twobody}.

Figure~\ref{fig:core_collapse_Rc_lag} shows the evolution of the core radius (top panel), $\Rc$, defined by
\begin{equation}
      \label{eq:Rc}
      \Rc^2 = \frac{ \sum_{i=1}^N \rho(\br_i)^2 \,r_i^2 }{ \sum_{i=1}^N \rho(\br_i)^2},
  \end{equation}
where $\br_i$ is the position of the $i^{\rm th}$ particle and $r_i$ its distance to the center of the SCF reference frame. We also display several Lagrange radii (bottom panel) and compare \krios\ predictions to measurements from $N$-body simulations.
We observe that core collapse is reached at $t_{\rc\rc}\sim 18 \,\trh$ (Fig.~\ref{fig:core_collapse_Rc_lag}) -- a value that weakly depends on the run parameters. This is similar to $N$-body measurements and Fokker--Planck predictions \citep{Takahashi1995}. The core collapse time is weakly dependent on the number of neighbors used to compute the local-mass density. The core collapse profile closely matches the direct $N$-body result through the late stages of core collapse, at which point a greater number of basis elements might be necessary to properly fit the mean field. We also compare to results obtained with \raga\ \citep{Vasiliev2015}, in which the 2-body relaxation in handled using local velocity diffusion coefficients instead of effective hyperbolic encounters. We obtain results reasonably consistent with our \krios\ run.

The mass density profile of a stellar system is well known to converge to a power law at small radii as the cluster approaches core collapse.  We show in Figure~\ref{fig:core_collapse_density} the stellar mass density at initial time (in blue), at an intermediary timestep (in green) and at the last snapshot of the core collapse run 
(in orange). The mass density converges to a power law in the inner regions at late times, with the 3D density between between $r=0.05\,\rm{HU}$ and 
$r=0.5\,\rm{HU}$ 
following a $\rho(\br) \propto r^{-2.35}$, which is very close to the $\rho(\br) \propto r^{-2.23}$ value predicted by Fokker--Planck models \citep{Takahashi1995}.  The outer regions follow a $\rho(\br) \propto r^{-4.03}$ power law (from $r=10\,\rm{HU}$ to 
$r=100\,\rm{HU}$), showing clear convergence towards the Fokker-Planck prediction of $\rho(\br) \propto r^{-3.5}$.   
It is likely that increasing the accuracy of the mode selection (and possibly the number of radial basis elements) would even further improve this fit at late 
times.

To verify the validity of the \krios\ code, we show in Figure~\ref{fig:IOM} the time evolution of the fractional energy change, $|\Delta E/E|$, and that of the absolute error in angular momenta, $|\Delta L|$ and $|\Delta\Lz|$.
On the one hand, energy conservation (bottom panel) remains satisfactory during relaxation due to the \stod\ prescription and stays below 1\% relative error at the end of the run. On the other hand, although angular momentum is only approximately conserved during 2-body encounters and suffers from the \stod\ prescription, it still remains relatively low -- below $|\delta \bL| \leq 0.06\,\rHU$  -- as long as we use the  density pairing scheme (see Appendix~\ref{app:iom_radial_v_density}). 

\begin{figure}[t] 
    \centering
    \includegraphics[width=0.95\columnwidth]{ 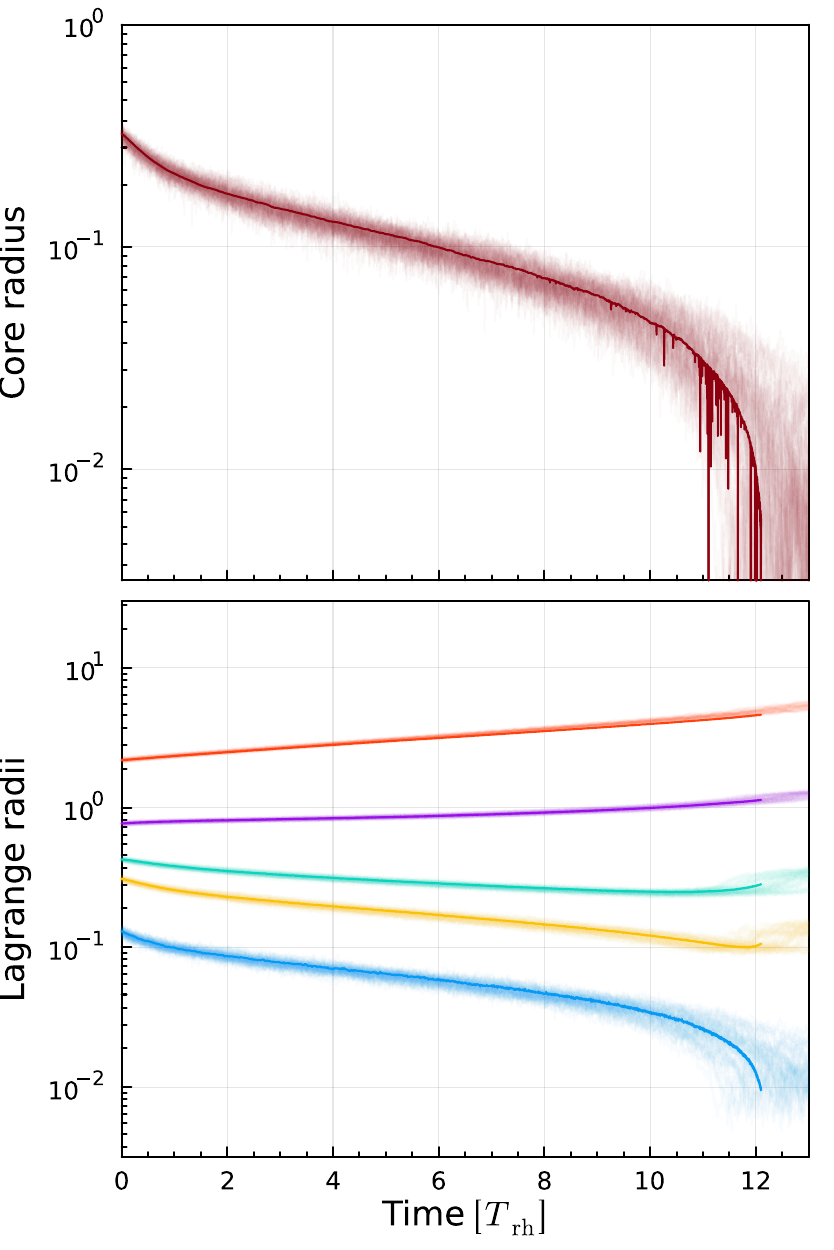}
   \caption{Same as Figure~\ref{fig:core_collapse_Rc_lag}, but instead starting from a $q=-6$ tangentially anisotropic cluster. As in the isotropic case, the SCF approach is consistent with measurements made in $N$-body simulations over the course of the cluster's relaxation up to the deeper stages of core collapse. Changes in the integrals of motion are kept below $< 1\%$ relative error in energy and $|\delta \bL| \leq 0.004 \,\rHU$. }
   \label{fig:core_collapse_Rc_lag_q_-6}
 \end{figure}

\begin{figure}[t] 
    \centering
    \includegraphics[width=0.95 \columnwidth]{ 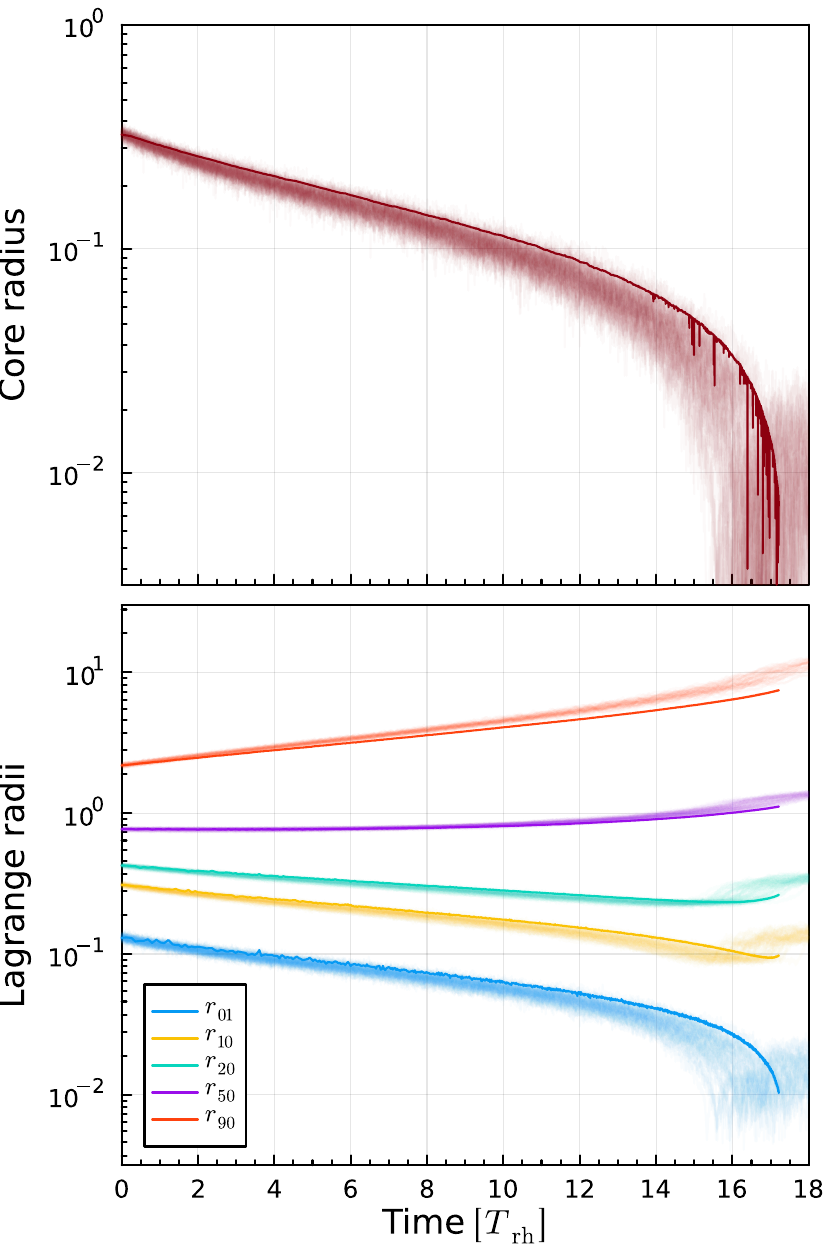}
   \caption{Same as in Figure~\ref{fig:core_collapse_Rc_lag}, instead starting from an isotropic rotating Plummer sphere with LBD parameter $\alpha_{\rr}=0.5$.  As in the non-rotating case, the SCF approach is consistent with measurements made in $N$-body simulations over the course of the cluster's relaxation up to the deeper stages of core collapse. Changes in the integrals of motion are once more kept below $< 1\%$ relative error in energy and $|\delta \bL| \leq 0.03\,\rHU$.
      }
   \label{fig:core_collapse_Rc_lag_rot}
 \end{figure}
\subsection{Relaxation of an anisotropic cluster}
\label{subsec:cc_q_-6}

We complement our testing of \krios\ by repeating the experiment described in section~\ref{subsec:CC_q_0} in the case of a tangentially anisotropic Plummer sphere with parameter $q=-6$ (see Appendix~\ref{app:plummer_cluster}) with $N=10^6$ stars.

Figure~\ref{fig:core_collapse_Rc_lag_q_-6} shows the long-term evolution of the core radius and the Lagrange radii obtained by running  \krios, and compares them against measurements made with $N$-body simulations. We find that \krios\ is able to  reproduce the expected  long-term $N$-body relaxation with satisfying accuracy, including the latest stages of core collapse. We report a conservation of energy well below 1\%  percent relative error at the end of the run, as well as a conservation of the angular momentum below an absolute deviation below $|\delta \bL| \leq 0.004\,\rHU$.

\subsection{Relaxation of a rotating cluster}
\label{subsec:relaxation_q_0_alpha_0.5}

Finally, we wish to probe the validity of \krios\ in the regime with with non-zero internal angular momentum. For that purpose, we introduce rotation to the system by following  Lynden-Bell's daemon (LBD) prescription \citep{LyndenBell1960}, which consists on changing a fraction $\alpha_{\rr} \in[0,1]$ of retrograte stars (with $\Lz<0$) into prograde stars (with $\Lz>0)$. 
In this section, we consider the evolution of an isotropic Plummer sphere with parameter $\alpha_{\rr}=1/2$. 

As in the previous sections, we first perform a comparative study  of the long-term evolution of the core radius and the Lagrange radii against a sample of 50 realizations of $N$-body runs. Figure~\ref{fig:core_collapse_Rc_lag_rot} shows that, once more, \krios\ is able to reproduce the $N$-body measurements to a satisfactory degree. We report a conservation of energy well below 1\% percent relative error at the end of the run, as well as a conservation of the angular momentum below an absolute deviation of $|\delta \bL| \leq 0.03\,\rHU$  -- corresponding to a 25\% percent relative error. 

\begin{figure} 
    \centering
    \includegraphics[width=0.99\columnwidth]{ 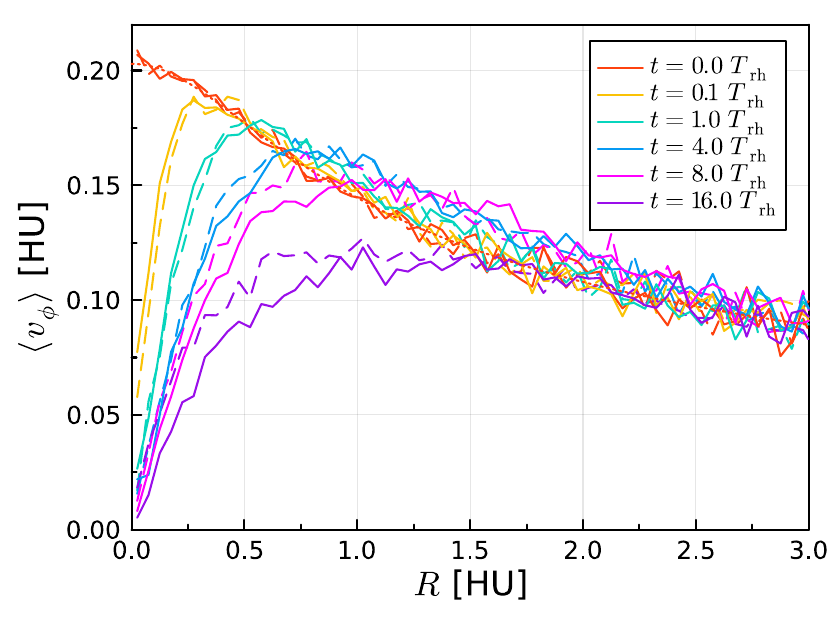}
   \caption{  Evolution of the rotation curve, $\langle v_{\phi}\rangle(R)$, along the relaxation of an isotropic, rotating cluster with $\alpha_{\rr}=1/2$. We compare \krios\ runs with $N=10^5$ particles averaged over 5 realizations (solid lines) against $N$-body runs with $N=10^4$ particles averaged over 50 realizations (dashed lines). The red dotted line is the theoretical prediction at $t=0$. As the cluster undergoes secular relaxation, its internal angular momentum is slowly reorganized. This phenomenon is captured both in $N$-body measurements and in \krios\ measurements.
   }
   \label{fig:vphi_rot}
 \end{figure}

Second, we can keep track of the exchange of angular momentum within the cluster by measuring the evolution of the the rotation curve, $\langle v_{\phi}\rangle (R)$, along core collapse. Figure~\ref{fig:vphi_rot} compares measurements made using a sample of \krios\ runs and a sample of $N$-body runs, and shows that \krios\ is consistent -- with the expected average $N$-body behavior. The deviations observed with $N$-body measurements -- which increase as time goes by -- might be the result of increasingly poor angular momentum conservation (reaching about 25\% relative error at core collapse, due to both the non-conservation of angular momentum in the 2-body encounter and the \stod\ prescription). In fact, the radial pairing prescription displays much worse angular momentum conservation -- reaching more than 100\% relative error at the end of the run -- and fails to even reproduce the $N$-body measurement at $t=0.1 \,\trh$. 

\subsection{Timing of the code}

\begin{figure}[tb] 
\hspace{-5mm}
    \centering    \includegraphics[width=0.99\columnwidth]{ 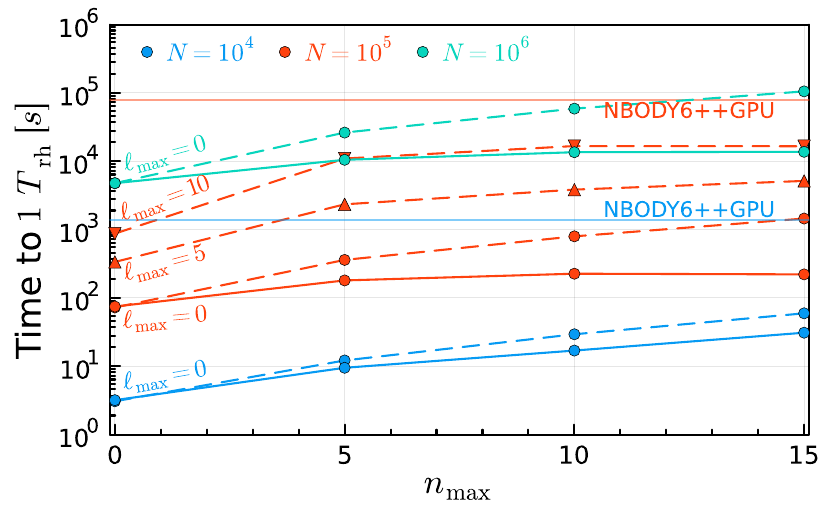}
   \caption{Impact of the \krios\ parameters on the time required to reach $1 \,\trh$, over 128 CPUs, where we used 25 NNs for the local density estimates (the impact of the number of NNs on runtime is negligible compared to the other parameters) and 1 sample per dynamical time. Solid lines show timings with basis filtering, while dashed lines show timings without basis filtering. The duration scales linearly with $n_{\max}$ (without basis filtering) and almost-linearly with $N$, while basis filtering speeds up force calculation. The effect of filtering sometimes results in the calculation of a single basis element of the SCF expansion thanks to our use of the optimized \citetalias{Zhao1996} basis family.
      }
   \label{fig:timing_chc}
 \end{figure}

We complete this study by measuring how the \krios\ code scales with respect to the number of basis elements ($n_{\max}$, $\ell_{\max}$), the number of particles $N$, and by showing the difference between a non-filtered basis and a filtered basis. 
Figure~\ref{fig:timing_chc} summarizes these measurements and provides a comparison with \texttt{NBODY6++GPU}. 

As we may have expected from the SCF expansion, the code scales linearly with $n_{\max}$ (Equation~\ref{eq:SCF_expansion}) and quasi-linearly with $N$ (due to basis coefficients calculation, see Equation~\ref{eq:ap_from_N}, and particle sorting). Basis filtering (in dashed lines) greatly reduces the computation time compared to unfiltered basis (in solid lines). In fact, this speed-up is improved further by our use of the optimized \citetalias{Zhao1996} basis elements. In some cases, the ${(\alpha,b)}$ choice returns a single basis element during the early stages of core collapse. Finally, \krios\ is faster than the GPU-accelerated \texttt{NBODY6++GPU} code by at least an order of magnitude, rendering it suitable for the study of large-$N$ systems and the exploration of large parameter spaces.

 \section{Conclusions and perspectives}
\label{sec:conclusions}

 \subsection{Conclusions}

In this paper, we presented a new code, \krios, that combines a 3D collisionless evolution using a self-consistent field method with the two-body collisional dynamics of H\'{e}non's method. To do so, we fit a biorthogonal set of potential and mass-density basis functions to the cluster's current configuration at each system timestep, allowing for the efficient calculation of the mean field forces. Collisional effects were taken into account through pairwise gravitational scattering events of nearby particles, which were paired following a local density-pairing method. 

Firstly, we were able to reproduce the radial-orbit instability, a collisionless property of sufficiently radially anisotropic Plummer spheres. This instability quickly deforms the Plummer sphere into a triaxial cluster, which requires to break spherical symmetry. Therefore, such a reproduction is an excellent confirmation of our SCF usage. \krios\ in its collisionless form was able to reproduce both the initial linear growth regime measured in $N$-body simulations, \texttt{EXP} \citep{EXP2022} runs, and linear response theory \citep{Petersen2024}. It also resolved the nonlinear regime following linear growth -- including amplitude saturation and slow oscillatory decay. 

Secondly, we were able to integrate a family of $N=10^6$ clusters with varying anisotropy and angular momentum content to core collapse very efficiently compared to direct the GPU-accelerated $N$-body code 
\texttt{NBODY6++GPU} \citep{Wang2015}, while both conserving energy throughout the run with less than 1\%  relative error and keeping the angular momentum drift relatively low. Core collapse was shown to occur around 
$18 \,\trh$ for the isotropic Plummer sphere,  $12.2 \,\trh$  for the $q=-6$ Plummer sphere and $17.2 \,\trh$ for the rotating, isotropic Plummer sphere, which are consistent with $N$-body simulations (and Fokker--Planck models in the isotropic case). These values are weakly dependent on the run parameters -- e.g., the number of basis elements. In addition, the core radius and the Lagrange radii we measured in our \krios\ run appeared to be consistent with that of $N$-body realizations, up to a small divergence occurring at later times which we may attribute to an insufficient  number of basis elements. 
We were also able to measure the cusp of the density near core collapse, and observed its convergence to the $\rho(\br) \propto r^{-2.23}$ cusp predicted by \citet{Cohn1980} and measured by \citet{Takahashi1995}.  Furthermore, the \citetalias{Zhao1996} basis elements were shown to be able to fit the potential-density pair of the cluster with great accuracy during the entire core collapse process, and its parameters $\alpha$ and $b$ appeared to be good qualitative proxies to measure the degree to which a cuspy center had developed in the core and the size of the cluster. We also showed that \krios\ could reproduce the evolution of the rotation curve, $\langle v_{\phi}\rangle (R)$, in the case of a rotating LBD Plummer sphere, though the accuracy was limited by the approximate conservation of angular momentum in our 2-body relaxation scheme.

 \subsection{Perspectives}

One of the primary benefits of breaking spherical symmetry in a GC dynamics code is the ability to more realistically model the interactions between the cluster and its environment. \krios\ is well-suited modeling GCs subject to external tidal fields from a host potential, especially on orbits where the $\ell>0$ basis elements become more pronounced and tidal debris is produced. The $\mathcal{O}(N\log N)$ complexity makes model testing more numerically efficient than direct $N$-body simulations, whereas the complete 6D information made available for all stream and progenitor stars makes it more comprehensive than existing particle-spray codes. In the second paper of this series, we will describe the supporting infrastructure for \krios\ runs of this kind, such as the implementation of an external tidal field and the response of the cluster, as well as the production of stellar streams from the escaping stars \citep{Cook2025}. 

Because \krios\ can evolve clusters on a star-by-star basis, future work will also focus on the development of additional stellar physics, such as stellar evolution for stars and binaries (as is done in other direct $N$-body and Monte Carlo codes).  We also plan to consider strong encounters between nearby particles, such as binary star encounters and dynamical binary formation.  This can be done by sampling from the distribution of possible encounters given a list of neighbors, then performing small-$N$ scattering experiments with a direct $N$-body code \cite[similar to Monte Carlo codes, e.g.,][\S 2.2]{Rodriguez2023}.  However, by tracking the 6D spatial information of these encounters, \krios\ will be able to track the orientation and angular momentum of these binaries as well, something not currently possible in current Monte Carlo codes.  This will allow for a detailed study of the long term dynamical evolution of binaries in star clusters, from stellar mass binaries to supermassive black holes.

\section*{Data Distribution}

The data used to construct each of these figures will be made available upon request to the authors.

\section*{Acknowledgments}

We thank our referee, Douglas Heggie, for his
remarks, which greatly contributed to the improvement of this
paper.   We are also grateful to Eugene Vasiliev and Martin Weinberg for numerous suggestions during the completion of this work. We also thank Eugene Vasiliev for providing us with the \raga\ run data used in this paper. This work is partially supported by the National Science Foundation under Grants No. AST-2310362 to the University of North Carolina and No. PHY-2309135 to the Kavli Institute for Theoretical Physics (KITP), by NASA ATP Grant 80NSSC24K0687, and by the grant \href{https://www.secular-evolution.org}{\emph{SEGAL}} ANR-19-CE31-0017
of the French Agence Nationale de la Recherche. CR also acknowledges
support from an Alfred P. Sloan Research Fellowship, and a
David and Lucile Packard Foundation Fellowship.  BTC was partially funded by the North Carolina Space Grant's Graduate Research Fellowship.
MSP is supported by a UKRI Stephen Hawking Fellowship.
We would like to thank the University of North Carolina at Chapel Hill and the Research Computing group for providing computational resources and support that have contributed to these research results.
We thank St\'ephane Rouberol for the smooth running of the
Infinity cluster of the Institute of Astrophysics of Paris, where the $N$-body runs were performed.

\appendix

\counterwithin{figure}{section}
\counterwithin{table}{section}
\renewcommand\thefigure{\thesection.\arabic{figure}}

\section{The SCF approach}
\label{app:scf}

\subsection{The Zhao SCF expansion}
\label{app:coeffs_SCF}

We consider the basis expansion
\begin{subequations}
\begin{align}
\psi(\br)&= \sum_p a_p\, \psi^{(p)}(\br),\\
\rho(\br)&= \sum_p a_p\, \rho^{(p)}(\br).
\end{align}
\end{subequations}
Taking inspiration from the \citetalias{CB73} basis elements \citep{CB73, Fouvry2021}, we write the \citetalias{Zhao1996}
 basis elements in the form
\begin{subequations}
\label{eq:SCF_radial_harmonics}
\begin{align}
\psi^{(p)}(\br) &= Y_{\ell}^m(\vartheta,\phi)\,U_n^{\ell}(r),\\
\rho^{(p)}(\br) &= Y_{\ell}^m(\vartheta,\phi)\,D_n^{\ell}(r),
\end{align}
\end{subequations}
where
\begin{subequations}
\begin{align}
U_n^{\ell}(r)&\!=\!-\sqrt{\frac{G}{b}}\frac{\sqrt{4\pi}}{\sqrt{N_n^{\ell}}}\frac{(r/b)^{\ell}\, C_{n}^{w}(\xi)}{[1+(r/b)^{1/\alpha}]^{\alpha+2 \ell \alpha}} ,\label{eq:Unl_Zhao}\\
D_n^{\ell}(r)&\!=\! \frac{1}{\sqrt{G}\,b^{5/2}}\frac{\sqrt{4\pi} K_n^{\ell} }{\sqrt{N_n^{\ell}}}\,\frac{(r/b)^{\ell-2+1/\alpha}  C_{n}^{w}(\xi)}{[1\!+\!(r/b)^{1/\alpha}]^{2+\alpha+2 \alpha \ell}},
\end{align}
\end{subequations}
with
\begin{subequations}
\begin{align}
w&=(2\ell+1)\alpha + 1/2, \\
\xi&=\frac{(r/b)^{1/\alpha}-1}{(r/b)^{1/\alpha}+1},
\end{align}
\end{subequations}
and $C_{n}^{w}(\xi)$ are the Gegenbauer polynomials.
The numerical prefactors are given by
\begin{subequations}
\begin{align}
K_n^{\ell} &= \frac{4(n+w)^2-1}{16\pi \alpha^2}, \\
\frac{1}{N_n^{\ell} }&=\frac{2^{4w+1}\alpha(n+w)}{\pi[4(n+w)^2-1]}\frac{n! \Gamma(w)^2}{\Gamma(2w+n)},
\end{align}
\end{subequations}
where $\Gamma$ is the Gamma function. 
The special case $\alpha=1/2$ corresponds to the \citetalias{CB73} basis family with the shift $n\rightarrow n+1$, while $\alpha=1$ corresponds to the biorthogonal basis set given by \citet{HO1992}.

We compute the basis expansion coefficients by taking advantage of biorthogonality
\begin{align}
\int \rd \br \rho(\br) \psi^{(p)}(\br)^{*}&\!=\! \sum_q a_q\, \int \rd \br\rho^{(q)}(\br) \psi^{(p)}(\br)^{*}, \\
-\int \rd \br \rho(\br) \psi^{(p)}(\br)^{*} &\!=\! a_{p}.
\end{align}
Now, consider a system of $N$ particles, with masses $m_k$ and positions $\br_k$. Its instantaneous mass density function is given by
\begin{equation}
\rho(\br) = \sum_k m_k \deltaD^{(3)}(\br-\br_k).
\end{equation}
This yields the SCF coefficients
\begin{align}
\label{eq:ap_def}
a_p &=-\sum_k m_k \int \rd \br  \deltaD^{(3)}(\br-\br_k) \psi^{(p)}(\br)^{*}\\
&=-\sum_k m_k  \psi^{(p)}(\br_k)^{*}\notag.
\end{align}
Therefore,
\begin{align}
a_{\ell m n} 
&=-\sum_k m_k  U_{n}^{ \ell}(r_k) Y_{\ell}^m (\vartheta_k,\phi_k)^{*} .
\end{align}
In addition, using the relation ${Y_{\ell}^{-m}\!=\!(-1)^m (Y_{\ell}^m)^{*}}$, we obtain the formula
\begin{align}
a_{\ell -m n} 
&= (-1)^m a_{\ell m n}^{*}.
\end{align}

\subsection{Dealing with self-gravity}
\label{subsec:self_gravity}

The gradient of the SCF contains contributions from all bound particles, while in direct $N$-body the forces due to the other $N-1$ are considered explicitly. This incongruity introduces a self-gravity term in \krios, which we need to remove for bound particles. 
The true potential felt by a bound particle at $\br_k$ reads
\begin{align}
    \tilde{\psi}(\br_k) = \sum_{j \neq k} \,m_j \,U(|\br_j - \br_k|),
\end{align}
where $U(r)=-G /r$ is the Newtonian interaction potential. Using the relation \citep{FouvryCourses}
\begin{align}
    U(|\br-\br'|) = - \sum_p \,\psi^{(p)}(\br')^{*}\,\psi^{(p)}(\br),
\end{align}
the potential can be written as
\begin{align}
    \tilde{\psi}(\br_k) &=- \sum_{j \neq k} \,m_j \,\sum_p \,\psi^{(p)}(\br_j)^{*}\,\psi^{(p)}(\br_k)\\
    &=  \sum_p \bigg[a_p + m_k\,\psi^{(p)}(\br_k)^{*}\bigg]\,\psi^{(p)}(\br_k),\notag
\end{align}
where $a_p$ is given by Equation~\eqref{eq:ap_def}. It follows that
\begin{align}
    \tilde{\psi}(\br_k) 
     &=  \psi(\br_k) + \overbrace{m_k \sum_p  |\psi^{(p)}(\br_k)|^2}^{\psi_{\mathrm{sg},k}},
\end{align}
where $\psi_{\mathrm{sg},k}$ is the self-gravity potential felt by the particle $k$ and $ \psi(\br_k)$ is the SCF expansion of the (mean field) potential.
In contrast, unbound particles in \krios\ only feel the potential of the cluster, that is, generated by the bound particles. As such, we do not need to consider a self-gravity contribution in addition to the SCF expansion of the potential.

\subsection{Optimization of $a_{000}$ against $b$ and $\alpha$}
\label{app:opti_Zhao}

We have that
\begin{align}
a_{000} 
&=-\sum_k m_k  U_{0}^{0}(r_k) Y_{0}^0 (\vartheta_k,\phi_k)^{*} \\
&= -\frac{1}{\sqrt{4\pi}} \sum_k m_k  U_{0}^{0}(r_k),\notag
\end{align}
where
\begin{align}
U_0^{0}(r)&=-\sqrt{\frac{G}{b}}\frac{\sqrt{4\pi}}{\sqrt{N_0^{0}}}\frac{ C_{0}^{w}(\xi)}{[1+(r/b)^{1/\alpha}]^{\alpha}}\\
&=-\sqrt{\frac{G}{b}}\frac{\sqrt{4\pi}}{\sqrt{N_0^{0}}}\frac{ 1}{[1+(r/b)^{1/\alpha}]^{\alpha}},\notag
\end{align}
with $w=\alpha + 1/2$ and 
\begin{align}
\frac{1}{\sqrt{N_0^{0} }}
&=\frac{\sqrt{\Gamma(3+2\alpha)}}{\sqrt{2}\,\Gamma(2+\alpha)}.
\end{align}
Therefore,
\begin{align}
a_{000} 
&=   \sqrt{\frac{G\, \Gamma(3+2 \alpha)}{2 b \,\Gamma(2+\alpha)^2}}\sum_k\ \frac{ m_k}{[1+(r_k/b)^{1/\alpha}]^{\alpha}}.
\end{align}
We then optimize $a_{000}$ with respect to $\alpha$ and $\ln b$ through gradient ascent.

\subsection{Dynamical quantities via SCF}
\label{app:expression_SCF}

Let us consider the enclosed mass (for the bound stars) 
\begin{align}
M(r) =  \int_{0}^{r} \rd r\, r^2 \int \rd^2 \mathrm{S} \, \rho(\br),
\end{align}
where $ \rd^2 \mathrm{S} $ is the volume element of the sphere. Using the density basis, we may write
\begin{align}
M(r) \!=\! \sum_p a_p\! \int_{0}^{r}\! \rd r'\, r^{\prime 2}  D_{n}^{\ell}(r') \int \rd^2 \mathrm{S} \, Y_{\ell}^m(\vartheta,\phi).
\end{align}
Finally, since $Y_{0}^{0} = 1/\sqrt{4\pi}$, we can use the orthogonality of spherical harmonics, which yields
\begin{align}
M(r) &= \sqrt{4\pi} \sum_{\ell m n} a_{\ell m n} \int_{0}^{r} \rd r'\, r^{\prime 2}  D_{n}^{\ell}(r') \delta_{\ell}^{0} \delta_m^{0}\\
&= \sqrt{4\pi} \sum_{ n} a_{0 0 n} \int_{0}^{r}  \rd r'\, r^{\prime 2}  D_{n}^{0}(r') \notag.
\end{align}
The enclosed mass 
can be expressed in terms of the reduced radius variable, ${\xi\!=\!\xi(r)}$, as 
\begin{align}
M(r)  
\!=\! \frac{2   \pi \alpha }{2^{\alpha}}\sqrt{b \over G} \sum_{n} \frac{a_{00n} K_{n}^{0} }{ \sqrt{N_{n}^{0}}}\hspace*{-1mm} \int_{-1}^{\xi} \hspace*{-3mm}\rd \xi' \left(1+\xi'\right)^{\alpha} C_{n}^{\alpha+\tfrac{1}{2}}(\xi'),
\end{align}

\section{Stod\'o\L{}kiewicz prescription}
\label{app:stod_prescription}

The energy of particle $k$ at a given time reads
\begin{equation}
    E(t)=\frac{1}{2}m_k \dot{\br}(t)^2+ m_k \,\psi(\br_k[t],t).
\end{equation}
Taking the derivative and applying Newton's equations yields
\begin{equation}
    \dot{E}(t)= m_k \,\frac{\p \psi}{\p t}(\br_k[t],t).
\end{equation}
We may therefore express the energy  change experienced by particle $k$  between the time $t_{i}$ and $t_e$ as \citep{stodolkiewicz1982} 
\begin{align}
\label{eq:dE_stod_exact}
\Delta E=m_k\int_{t_i}^{t_e} \hspace*{-3mm} \rd t \,\frac{\p \psi}{\p t}(\br_k[t],t).
\end{align}
This integral may be estimated using the coarsest trapezoidal scheme
\begin{align}
\Delta E&\simeq \frac{m_k \Delta  t}{2}\bigg[ \frac{\p \psi}{\p t}(\br_k[t_i],t_i)+\frac{\p \psi}{\p t}(\br_k[t_e],t_e)\bigg]\\
&\simeq\frac{m_k}{2}\bigg[  \psi(\br_k[t_i],t_e)-\psi(\br_k[t_i],t_i)\notag\\
&\hspace{8mm}+   \psi (\br_k[t_e],t_e)- \psi (\br_k[t_e],t_i)\bigg]\notag,
\end{align}
where we used finite differences to estimate the partial time derivatives, with $\Delta  t=t_e-t_i$. This yields Equations~\eqref{eq:stod_eq}.

We may be tempted to obtain a better approximation of the energy change given in Equation~\eqref{eq:dE_stod_exact} by using the sampling nodes described in Section~\ref{subsec:integrator}. However, we cannot use this method because $\psi$ is not updated between the start of the timestep $t_i$ and its end $t_e$, making it impossible to obtain realistic estimations of $\p \psi/\p t$ in between timesteps.

\section{Integrator}
\label{app:integrator}

We use the GSL ODE solver \citep{gough2009} to evolve forward each particle's phase space $\brw$ one system timestep, setting up a set of six first-order equations
\begin{align}
\dot{\rw}_{i}(t) &= f_{\rw_{i}}\left(t, \brw[t]\right).  
\end{align}
The total gravitational potential $\psi$ is a sum over all relevant contributions.  In the case of an  isolated cluster, ${\psi=\psi_{\rm SCF}=\sum_{p}a_{p}\psi^{(p)}(\br)}$, where we use the \citetalias{Zhao1996} basis elements to expand the cluster potential in the SCF frame. In future work, the host potential $\psi_{\rm host}$ can be  directly added to the potential.

In order to obtain the Cartesian components $f_{v_{x_{i}}}$, we need to evaluate the accelerations resulting from the cluster potential ${\bf a} = -\nabla\psi$. Due to the SCF expansion of the potential, it follows that we merely need to compute the acceleration associated with the potential basis elements of the \citetalias{Zhao1996} SCF expansion. These accelerations are easily expressed using the spherical coordinates attached to the SCF frame \begin{subequations}
\label{eq:accelerations_SCF}
\begin{align}
f_{v_{r}} &= -\sum_{p} a_{p} \, Y_{\ell}^{m}(\vartheta,\phi) {\rd U_{n}^{\ell}(r) \over \rd r}, \\
f_{v_{\vartheta}}  &= -\sum_{p} a_{p} \, {\partial Y_{\ell}^{m}(\vartheta,\phi) \over \partial \vartheta} \frac{U_{n}^{\ell}(r)}{r}, \\
f_{v_{\phi}} &= -\sum_{p} a_{p} \, ({\rm i} m) \frac{Y_{\ell}^{m}(\vartheta,\phi)\,U_{n}^{\ell}(r)}{r\sin\vartheta} .
\end{align}
\end{subequations}
We may then convert these components into Cartesian coordinates by applying the appropriate rotation matrix:
\begin{align}
\begin{bmatrix}
f_{v_{x}} \\f_{v_{y}} \\f_{v_{z}} 
\end{bmatrix} &\!=\! 
\begin{bmatrix}
    \sin\vartheta\cos\phi & \cos\vartheta\cos\phi & -\sin\phi \\ 
    \sin\vartheta\sin\phi & \cos\vartheta\sin\phi & \cos\phi \\ 
    \cos\vartheta & -\sin\vartheta & 0
\end{bmatrix}\!
\begin{bmatrix}
f_{v_{r}} \\f_{v_{\vartheta}} \\f_{v_{\phi}} 
\end{bmatrix}.\!\!
\end{align}
The Jacobian associated with the cluster's field may be necessary for implicit integrators. To that end, we provide an expression of the Cartesian derivatives from the spherical derivatives
\begin{align}
    \hspace*{-0.3mm}\begin{bmatrix} \partial_{x}\\ \partial_{y}\\ \partial_{z}
    \end{bmatrix} &\!=\! 
    \begingroup
\renewcommand*{\arraystretch}{1.5}
    \begin{bmatrix} 
    \sin\vartheta\cos\phi & \displaystyle{\cos\vartheta\cos\phi\over r} & -\displaystyle{\sin\phi\over r\sin\vartheta} \\ 
    \sin\vartheta\sin\phi & \displaystyle{\cos\vartheta\sin\phi\over r} & \displaystyle{\cos\phi\over r\sin\vartheta}
    \\ \cos\vartheta& -\displaystyle{\sin\vartheta\over r} & 0 \end{bmatrix} 
    \endgroup
    \hspace*{-1.5mm}
    \begin{bmatrix} \partial_{r}\\ \partial_{\vartheta}\\ \partial_{\phi}
    \end{bmatrix}\!.
\end{align}
It follows that the Jacobian associated with the cluster's field may be written in the form
\begin{subequations}
\begin{align}
f_{v_{r}} &\!=\! -\!\sum_{p} a_{p} \, T_{3}(p ), \\
f_{v_{\vartheta}} &\!=\! -\!\sum_{p} a_{p} \, \frac{T_{2}(p )}{r}, \\
f_{v_{\phi}} &\!=\! -\! \sum_{p}a_{p} ({\rm i} m) \, \frac{T_{1}(p )}{r\sin\vartheta}, \\
\partial_{r}f_{v_{r}} &\!=\! -\!\sum_{p}a_{p} \, T_{6}(p ), \\
\partial_{r}f_{v_{\vartheta}} &\!=\! -\!\sum_{p} a_{p} \Big[\frac{T_{4}(p ) }{r}\!-\!\frac{ T_{2}(p )}{r^2}\Big], \\
\partial_{r}f_{v_{\phi}} & \!=\! -\! \sum_{p}a_{p}({\rm i} m)\Big[\frac{ T_{3}(p ) }{r\sin\vartheta}- \frac{ T_{1}(p )}{r^2\sin\vartheta}\Big],\! \\
\partial_{\vartheta}f_{v_{r}} &\!=\! -\!\sum_{p}a_{p} \, T_{4}(p ), \\
\partial_{\vartheta}f_{v_{\vartheta}} &\!=\! -\! \sum_{p}a_{p} \, \frac{T_{5}(p )}{r}, \\
\partial_{\vartheta}f_{v_{\phi}} &\!=\! -\! \sum_{p}a_{p}({\rm i} m) \, \Big[ \frac{T_{2}(p )}{r\sin\vartheta} - \frac{T_{1}(p )}{r \sin \vartheta \tan \vartheta}\Big], \\
\partial_{\phi}f_{v_{r}} &\!=\! -\!\sum_{p}a_{p}({\rm i} m) \, T_{3}(p ), \\
\partial_{\phi}f_{v_{\vartheta}} &\!=\! -\! \sum_{p}a_{p}({\rm i} m) \, \frac{T_{2}(p )}{r}, \\
\partial_{\phi}f_{v_{\phi}} &\!=\! -\! \sum_{p}a_{p}(-m^{2}) \, \frac{T_{1}(p )}{r\sin\vartheta},
\end{align}
\end{subequations}
where $p=(n,\ell,m)$ and we defined the jump tables ${\{T_{i}(p, \br)\}}$ as \begin{subequations}  
\begin{align}
T_{0}(p, \br) &= Y_{\ell}^{m}(\vartheta, \phi) \, D_{n}^{\ell}(r), \\
T_{1}(p, \br) &= Y_{\ell}^{m}(\vartheta, \phi) \, U_{n}^{\ell}(r), \\
T_{2}(p, \br) &= {\partial Y_{\ell}^{m}(\vartheta, \phi) \over \partial \vartheta} \, U_{n}^{\ell}(r), \\
T_{3}(p, \br) &= Y_{\ell}^{m}(\vartheta,\phi) \, {\rd U_{n}^{\ell}(r) \over \rd r}, \\
T_{4}(p, \br) &= {\partial Y_{\ell}^{m}(\vartheta, \phi) \over \partial \vartheta} \, {\rd U_{n}^{\ell}(r) \over \rd r}, \\
T_{5}(p, \br) &= {\partial^{2}Y_{\ell}^{m}(\vartheta,\phi) \over \partial \vartheta^{2}} \, U_{n}^{\ell}(r), \\
T_{6}(p, \br) &= Y_{\ell}^{m}(\vartheta,\phi) \, {\rd^{2} U_{n}^{\ell}(r) \over \rd r^{2}}.
\end{align}
\end{subequations}
These jump tables greatly improve the efficiency of the computation of the spherical harmonics and of the Gegenbauer polynomials. In turn, this greatly improves the efficiency of the calculation of the various ingredients needed to solve the ODE numerically.

\section{Amplitude of the $\ell^{\rm th}$ harmonic}
\label{app:amplitude_ell}

Let us consider an arbitrary stellar distribution, $\rho(\br)$, with its corresponding potential, $\psi(\br)$. We may expand these functions over a biorthonormal basis family (see Equations~\ref{eq:SCF_radial_harmonics}). It follows that
\begin{subequations}
    \begin{align}
        \rho(\br) &= \sum_{\ell m n} a_{\ell m n} \,D_n^{\ell}(r) Y_{\ell}^m(\vartheta,\phi),\\
        \psi(\br) &= \sum_{\ell m n} a_{\ell m n} \,U_n^{\ell}(r) Y_{\ell}^m(\vartheta,\phi),
    \end{align}
\end{subequations}
and where the SCF coefficients are given by
\begin{align}
    a_{\ell m n} = - \int \rd \br \, \rho(\br) U_n^{\ell}(r) Y_{\ell}^m(\vartheta,\phi)^{*}.
\end{align}
The harmonic amplitude is therefore given by
\begin{align}
    E_{\ell} = \sum_{n m} |a_{\ell m n}|^2.
\end{align}
Now, let us consider the rotation matrix $\brR$, which we apply to the stellar system. The rotated stellar distribution and potential are given by
\begin{align}
    \tilde{\rho}(\br) = \rho\big (\brR^{-1} \cdot \br \big) ;\quad
    \tilde{\psi}(\br) = \psi\big (\brR^{-1} \cdot \br \big),
\end{align}
where their SCF expansions are given by
\begin{subequations}
    \begin{align}
        \tilde{\rho}(\br) &= \sum_{\ell m n} \tilde{a}_{\ell m n} \,D_n^{\ell}(r) Y_{\ell}^m(\vartheta,\phi),\\
        \tilde{\psi}(\br) &= \sum_{\ell m n} \tilde{a}_{\ell m n} \,U_n^{\ell}(r) Y_{\ell}^m(\vartheta,\phi).
    \end{align}
\end{subequations}
It follows that their SCF coefficients are given by \begin{align}
    \tilde{a}_{\ell m n} 
    &= - \int \rd \br \, \tilde{\rho}(\br) U_n^{\ell}(r) Y_{\ell}^m(\vartheta,\phi)^{*}\\
    &= - \int \rd \br \, \rho\big (\brR^{-1} \cdot \br \big) U_n^{\ell}(r) Y_{\ell}^m(\vartheta,\phi)^{*}\notag\\
    &= - \int \rd \br' \, \rho(\br') U_n^{\ell}(r')   Y_{\ell}^m  (\brR[\vartheta',\phi'])^{*} \notag,
\end{align}
where we applied the change of variables $\br \mapsto \br'=\brR^{-1} \cdot \br$, used the orthogonality of the rotation matrix to compute the Jacobian $|\brR|=1$ and $\brR[\vartheta',\phi']$ stands for the new spherical angles obtained by applying the rotation matrix $\brR$ to the previous spherical angles $(\vartheta',\phi')$. Now, we may use the formula
\begin{align}
    Y_{\ell}^m  (\brR[\vartheta',\phi']) = \sum_{m'=-\ell}^{\ell} D_{mm'}^{(\ell)}(\brR)^{*} Y_{\ell}^{m'}(\vartheta',\phi'),
\end{align}
where $D_{mm'}^{(\ell)}(\brR)$ is an element of the Wigner matrix, $D_{\ell}(\brR)$, associated with the rotation matrix $\brR$ \citep[see, e.g.,][]{Edmonds1996}. This matrix is also orthogonal and satisfies the relation ${D_{\ell}(\brR)^{\dagger}D_{\ell}(\brR)=\brI}$. It follows that
\begin{align}
    \tilde{a}_{\ell m n} 
    &= - \sum_{m'} D_{mm'}^{(\ell)}(\brR) \int \rd \br' \, \rho(\br') U_n^{\ell}(r')   Y_{\ell}^m  (\vartheta',\phi')^{*} \notag\\
    &= \sum_{m'} D_{mm'}^{(\ell)}(\brR) \,\tilde{a}_{\ell m' n} .
\end{align}
Therefore, 
\begin{align}
 \tilde{E}_{\ell} &= \sum_{n,m} |\tilde{a}_{\ell m n}|^2=  \sum_{n,m} \tilde{a}_{\ell m n}^{*} \tilde{a}_{\ell m n}\\
 &=\sum_{n,m,m',m''} D_{mm'}^{(\ell)}(\brR)^{*} \, a_{\ell m' n}^{*} D_{mm''}^{(\ell)}(\brR) \, a_{\ell m'' n} \notag\\
 &=\sum_{n}\sum_{m',m''}\bigg[\sum_{m} D_{mm'}^{(\ell)}(\brR)^{*}D_{mm''}^{(\ell)}(\brR)\bigg]   a_{\ell m' n}^{*}  a_{\ell m'' n} \notag\\
  &=\sum_{n} \sum_{m',m''} \delta_{m'}^{m''} a_{\ell m' n}^{*} a_{\ell m'' n}=\sum_{n,m'}  a_{\ell m' n}^{*}  a_{\ell m' n} \notag\\
   &=\sum_{n,m} | a_{\ell m n}|^2 = E_{\ell}\notag,
\end{align}
which proves that $E_{\ell}$ is independent of $\brR$.

\section{The Plummer cluster}
\label{app:plummer_cluster}

The Plummer potential reads
\begin{align}
    \psi(r) = - \frac{G M}{\sqrt{b^2 + r^2}},
\end{align}
where $M$ is the total mass of the cluster and $b$ the Plummer scale length. Using Poisson equation, we can recover its associated mass density
\begin{align}
    \rho(r) = \frac{3 M}{4 \pi } \frac{b^2}{(b^2+r^2)^{5/2}},
\end{align}
which integrates to $M$. Let us define the characteristic quantities $E_0\!=\!-GM/b$ and $\Lo=\sqrt{GMb}$. Then, using Eddington inversion \citep{Eddington1916}, we can obtain the isotropic DF \citep{Dejonghe1987}
\begin{align}
    F(E)= \frac{M}{L_0^3} \frac{24 \sqrt{2}}{7 \pi^3} \tE^{7/2},
\end{align}
where we introduced the rescaled energy $\tE = E/\Eo$, such that  
$\tE>0$ correspond to the energy of a bound particle. For unbound particles with $\tE \leq 0$, the DF vanishes.

Wile only one isotropic DF exists for a given spherical system, there exists an infinite number of anisotropic models. In this paper, we consider the Osipkov-Merritt parameterization \citep{Osipkov1979,Merritt1985}. 
Letting $Q=\big[E+L^2/(2\rra^2)\big]/E_0$, the DF reads \citep{Petersen2024}
\begin{align}
    F(Q)=  \frac{M}{L_0^3}  \frac{24 \sqrt{2}}{7 \pi^3} Q^{7/2} \bigg[1 - \frac{b^2}{\rra^2}+\frac{7 b^2}{16 \rra^2 Q^2}\bigg],
\end{align}
for $Q>0$, where $\rra$ is the anisotropy radius, and vanishes for $Q \leq 0$. As is the case for all Osipkov-Merritt parameterizations \citep{Binney2008}, its associated anisotropy parameter computes to 
\begin{align}
    \beta(r) = 1-\frac{\sigma_{\rt}^2}{2 \sigma_{\rr}^2} =\frac{r^2}{r^2+\rra^2}.
\end{align}
Finally, we note that the $\rra=b$ case corresponds to the $q=2$ radial DF of the anisotropic family studied by \cite{Dejonghe1987}, given by \begin{equation}
\Ftot (E,L;q) = \frac{M}{\Lo^{3}}\frac{3 \, \Gamma (6 \!-\! q) \, \tE^{7/2-q}}{2(2\pi)^{5/2}} \, \mH_q \bigg(\frac{\tL^{2}}{2\tE}\bigg).
\label{eq:def_Fq}
\end{equation}
In Equation~\eqref{eq:def_Fq},
we introduced the rescaled angular momentum, $\tL=L/\Lo$.
We also introduced the function
\begin{equation}
\mathbb{H}_q(x) \!=\! \begin{cases}
\displaystyle{\frac{\!\HG(\half q,q \!-\!\tfrac{7}{2},1;x)}{\Gamma(\tfrac{9}{2} \!-\! q)} }, & {\mathrm{if}} \ x\leq1 ,
\\[2.0ex]
\displaystyle{\frac{\!\HG(\half q, \half q, \half (9 \!-\! q);1/x)}{\Gamma(1 \!-\! \half q)\Gamma(\half (9 \!-\! q))} } \frac{1}{x^{q/2}},&{\mathrm{if}}\ x \geq1 ,
\end{cases}
\end{equation}
where $\HG$ is the hypergeometric function. Its associated anisotropy parameter computes to 
\begin{align}
    \beta(r) = 1-\frac{\sigma_{\rt}^2}{2 \sigma_{\rr}^2} =\frac{q}{2}\frac{r^2}{r^2+b^2}.
\end{align}

\section{Conservation of angular momentum using the radial pairing scheme}
\label{app:iom_radial_v_density}

\begin{figure} 
    \centering
    \includegraphics[width=0.95 \columnwidth]{ 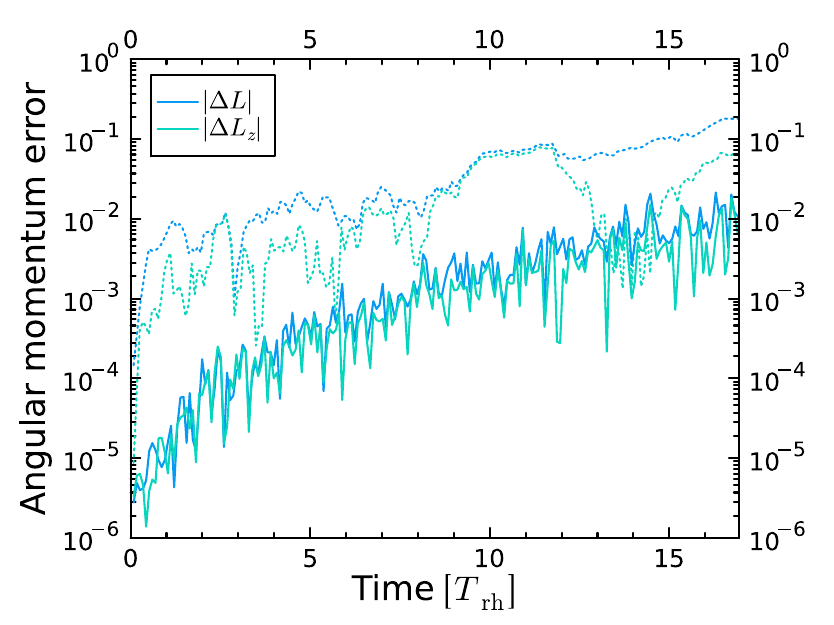}
   \caption{ Evolution of the change in angular momentum (in H{\'e}non units), $|\Delta L|, |\Delta\Lz|$, as in the bottom panel of Figure~\ref{fig:IOM} (non-rotating isotropic cluster), using the radial pairing scheme (dotted lines) and the local-density pairing scheme (solid lines) for an isotropic Plummer sphere. The latter method yields much better conservation, with an error in the angular momentum about 10 times lower than with  the radial pairing method.
      }
   \label{fig:IOM_radial_v_density}
 \end{figure}

The radial pairing scheme described in Section~\ref{subsec:particle_pairing} has the possibility of  pairing two particles in very different locations. It follows that: (i) the two interacting particles are not necessarily subject to the same mass and velocity DFs, with catastrophic consequences in the rotating case; (ii) angular momentum is ill-conserved during the effective hyperbolic encounter.\footnote{The pairing method has no impact on the conservation of energy, which depends on the magnitudes of the radius and velocities. This observation is confirmed in numerical experiments.} Figure~\ref{fig:IOM_radial_v_density} compares the conservation of angular momentum when using the radial pairing scheme and the local-density pairing scheme. It clearly shows that the local-density pairing method is much better at conserving angular momentum than the radial pairing one.

\end{document}